\documentclass[acmsmall,10pt,screen]{acmart}
\usepackage{ifpdf}
\ifpdf
\pdfoutput=1 % Force arXiv to process with pdflatex.
\fi
\usepackage{booktabs} % For formal tables
\usepackage{pgf-umlsd}

\usepackage[ruled]{algorithm2e} % For algorithms

\usepackage{listings}

\usepackage{color}
\usepackage{dot2texi}
\usepackage{tikz}
\usetikzlibrary{shapes,arrows}

\usepackage[inline]{enumitem}

\usepackage{glossaries}

\makeglossaries

\newglossaryentry{plug-in}{
  name={plug-in},
  description={
    is a Scala \lstinline{trait} designed to be instantiated by \href{https://javadoc.io/page/com.thoughtworks.feature/factory_2.11/latest/com/thoughtworks/feature/Factory.html}{\lstinline{Factory}}, along with other plug-ins
  }
}

\newglossaryentry{differentiable function}{
  name={differentiable function},
  description={
    is a Scala function that returns a \gls{differentiable expression}. It may represent a loss functions, a neural network or a subset of a neural network (e.g. a dense block in DenseNet\cite{iandola2014densenet})
  }
}

\newglossaryentry{differentiable expression}{
  name={differentiable expression},
  description={
    is a composable expression that supports operator overloading, whose type is \lstinline{DoubleLayer}, \lstinline{FloatLayer}, \lstinline{INDArrayLayer}, or other subtypes of \href{https://javadoc.io/page/com.thoughtworks.deeplearning/plugins-layers_2.11/latest/com/thoughtworks/deeplearning/plugins/Layers$Layer.html}{\lstinline{Layer}}. After a differentiable expression is built, it can perform \lstinline{forward} pass to create a differentiable \glspl{computational graph}.
  }
}

\newglossaryentry{trainable variable}{
  name={trainable variable},
  description={
    is a scalar or vector weight in a model, whose type is \lstinline{DoubleWeight}, \lstinline{FloatWeight}, \lstinline{INDArrayWeight}, or other subtypes of \href{https://javadoc.io/page/com.thoughtworks.deeplearning/plugins-weights_2.11/latest/com/thoughtworks/deeplearning/plugins/Weights$Weight.html}{\lstinline{Weight}}
  }  
}

\newglossaryentry{computational graph}{
  name={computational graph},
  description={
    is an asynchronous monadic data type that manages the life cycle of tapes, whose type is \href{https://javadoc.io/page/com.thoughtworks.raii/asynchronous_2.11/latest/com/thoughtworks/raii/asynchronous$$Do.html}{\lstinline{Do}}\lstinline{[}\href{https://javadoc.io/page/com.thoughtworks.deeplearning/deeplearning_2.11/latest/com/thoughtworks/deeplearning/DeepLearning$$Tape.html}{\lstinline{Tape[Data, Delta]}}\lstinline{]}
  }
}

\setcopyright{none}

\newsavebox{\IdrisDeepLearning}

\begin{document}

% Title portion
\title{Monadic Deep Learning}
%  \titlenote{This is a titlenote}
\subtitle{Performing monadic automatic differentiation in parallel}
%  \subtitlenote{Subtitle note} 

\author{Yang, Bo}
\orcid{0000-0003-2757-9115}
\affiliation{
  \institution{ThoughtWorks, Inc}
%   \streetaddress{104 Jamestown Rd}
%   \city{Williamsburg}
%   \state{VA}
%   \postcode{23185}
}
\email{atryyang@thoughtworks.com}
\author{Zhang, Zhihao}
\affiliation{%
  \institution{ThoughtWorks, Inc}
%   \streetaddress{104 Jamestown Rd}
%   \city{Williamsburg}
%   \state{VA}
%   \postcode{23185}
}
\email{zhazhang@thoughtworks.com}
\author{Marisa Kirisame} 
\affiliation{
 \institution{University of Washington}
 \city{Seattle} 
 \state{Washington}
 \country{USA}}
\email{lolisa@marisa.moe}
\author{Shi, Kai} 
\authornote{The corresponding author}
\affiliation{
  \institution{ThoughtWorks, Inc}
%  \city{Seattle} 
%  \state{Washington}
%  \country{USA}
}
\email{kshi@thoughtworks.com}

\begin{abstract}
The Java and Scala community has built a very successful big data ecosystem.
However, most of neural networks running on it are modeled in dynamically typed programming languages.
These dynamically typed deep learning frameworks treat neural networks as differentiable expressions that contain many \glspl{trainable variable},
and perform automatic differentiation on those expressions when training them.

Until 2019, none of the learning frameworks in statically typed languages provided the expressive power of traditional frameworks.
Their users are not able to use custom algorithms unless creating plenty of boilerplate code for hard-coded back-propagation.

We solved this problem in DeepLearning.scala 2. Our contributions are:

\begin{itemize}
  \item We discovered a novel approach to perform automatic differentiation in reverse mode for statically typed functions that contain multiple \glspl{trainable variable}, and can interoperate freely with the metalanguage.
  \item We designed a set of monads and monad transformers, which allow users to create monadic expressions that represent dynamic neural networks.
  \item Along with these monads, we provide some applicative functors, to perform multiple calculations in parallel.
\end{itemize}

With these features, users of DeepLearning.scala were able to create complex neural networks in an intuitive and concise way, and still maintain type safety.
\end{abstract}




\maketitle

% % The default list of authors is too long for headers.
\renewcommand{\shortauthors}{B. Yang et al.}

\begin{lrbox}{\IdrisDeepLearning}
\begin{lstlisting}[basicstyle=\footnotesize,language=Haskell]
interface DeepLearning Differentiable where
  Data : Type
  Delta : Type
  forward : Differentiable -> Do (Tape Data Delta)
\end{lstlisting}
\end{lrbox}

\section{Background and Introduction}
% \label{Background and Introduction}

Deep neural networks have become the state of the art on many tasks, such as computer vision, game playing, voice recognition and natural language translation.

A neural network is a computation model that transforms the input, into output, by repeated application of tensor operations (including matrix multiplication, tensor resizing, element-wise operations, such as max, +, etc). A ``deep'' neural network just indicates there are large amount of such transformations.

Additionally, a neural network has additional hidden inputs, called parameters or weights, and deep learning is the task of finding the best weights for a neural network, such that a predefined objective (called loss) is minimized.

Deep learning is mostly done using different variation of gradient descend: we calculate the first order derivative of the loss function with respect to the weight, and update the weight accordingly. The processed is called backpropagation.

Thus, backpropagation\cite{rumelhart1985learning} is the key feature in many deep learning frameworks. Combined with other optimization algorithms\cite{kingma2014adam, zeiler2012adadelta,duchi2011adaptive}, deep learning frameworks mutate the values of weights in neural networks during training, producing a model of knowledge learnt from training data.

Backpropagation can be considered as a specialized instance of Automatic Differentiation (AD)\cite{baydin2015automatic}. Many successful Python deep learning frameworks\cite{tokui2015chainer,google2017eager,paszke2017pytorch,neubig2017dynet} implement a common set of features of auto differentiation:

\begin{description}

  \item[Reverse mode] These deep learning frameworks perform reverse mode AD instead of forward mode, as forward mode AD does not scale well for deep neural networks.

  \item[Multiple trainable variable] Neural networks are composed of multiple layers. Each layer contains their own trainable variables. These deep learning frameworks are able to calculate the derivatives of all trainable variables at once for one training data batch.

  \item[Internal DSL\cite{fowler2010domain}] These deep learning frameworks are libraries that provide an Internal DSL, allowing users to create their differentiable functions in Python or Lua from similar expressions as creating ordinary non-differentiable functions. Since these frameworks do not require external language, models created by them can be easily integrated into a larger application simply with higher-level configurations\cite{chollet2015keras} or ETL (Extract, Transform and Load) process.

  % \item Deep learning framework using Dynamic Graph approach allow one to reuse whatever construct already available in the metalanguage (closure, object, list, tree, control flow, exception, etc, and preexisting function over them)

  % \item Deep Learning framework let you define weight, or trainable variable, and the framework can automatically do gradient descend over them, without users having to manually manage them.

\end{description}

% However, most deep learning framework are in Python, which is dynamically typed.
% Hence, it is not clear how to design the framework in a statically typed language, such that the DSL itself can enjoy the static type checking of the metalanguage, using technique such as final tagless or GADT.

However, how to archive those goals in statically typed library is still an open question. Previous solutions in statically type languages usually requires metaprogramming or compiler-time translation. In this paper, we present DeepLearning.scala, which is the first statically typed implementation that achieves all the above goals for deep neural networks. Our approach is based on some widely available functional programming constructs, thus can be ported to other statically typed programming languages.

\section{Basic Concepts}\label{concepts}

For example, suppose we are building a robot for answering questions in IQ test like this:

\begin{quote}
  What is the next number in this sequence:
    \begin{quote}
    3, 6, 9, ?
    \end{quote}
  The answer is 12.
\end{quote}

In DeepLearning.scala, the robot can be implemented as a \lstinline{guessNextNumber} function as following\footnote{The code examples from Listing~\ref{guessNextNumber} to Listing~\ref{predict_trained} do not contain necessary \lstinline{import} and configurations. For an executable model backed by ND4J\cite{skymind2017nd4j}, see \href{http://deeplearning.thoughtworks.school/demo/GettingStarted.html}{Getting Started documentation on DeepLearning.scala website}.}:

\begin{lstlisting}[float={h t b p},caption={The differentiable matrix multiplication implemented by \lstinline{map}/\lstinline{reduce}},label={guessNextNumber}]
// Weight initialization
val weights: Seq[DoubleWeight] = Stream.continually(DoubleWeight(math.random))
val bias: DoubleWeight = DoubleWeight(math.random)

def guessNextNumber(question: Seq[Double]): DoubleLayer = {
  // Perform a dot product between question and weights
  (question zip weights).map {
    case (element, weight) => element * weight
  }.reduce(_ + _) + bias
}
\end{lstlisting}

\lstinline{guessNextNumber} performs a dot product between question and weights by invoking higher-order functions \lstinline{map} and \lstinline{reduce}. 

Unlike\cite{chen2017typesafe}'s special tensor type, our tensor can be typed simply as \lstinline{Seq[Double]}, \lstinline{Seq[DoubleWeight]} or \lstinline{Seq[DoubleLayer]}.
\footnote{DeepLearning.scala users can use other representations of tensors:
\begin{enumerate*}
  \item For tensors with a statically typed shape, use \href{https://javadoc.io/page/com.chuusai/shapeless_2.11/latest/shapeless/Sized.html}{ \lstinline{shapeless.Sized}}. For example a \(10\times20\) two-dimensional tensor can be typed as \lstinline{Sized[Sized[Double, _10], _20]}. For differentiable tensors, replace vanilla \lstinline{Double} to \lstinline{DoubleWeight} or \lstinline{DoubleLayer}.
  \item For GPU-accelerated tensors, use \lstinline{INDArray}\cite{skymind2017nd4j}. For differentiable tensors, use \lstinline{INDArrayLayer} or \lstinline{INDArrayWeight} instead.
\end{enumerate*}
}

The return value of \lstinline{guessNextNumber}, along with temporary variables in \lstinline{guessNextNumber}, are \lstinline{DoubleLayer}s, which are \glspl{differentiable expression}.

The \lstinline{weights} and \lstinline{bias} contains some \lstinline{DoubleWeight} that are referenced by \lstinline{guessNextNumber}.
They must be initialized before executing the model. Those are \glspl{trainable variable}.

From the user's point of view, both \lstinline{DoubleLayer} and \lstinline{DoubleWeight} are opaque types similar to an ordinary \lstinline{scala.Double}. Most of the operators of \lstinline{scala.Double} are also available on \lstinline{DoubleLayer}  and \lstinline{DoubleWeight}, except those operators are differentiable. For now, the type signature of the multiplication operator as can be seen as in Listing~\ref{hypothetical differentiable types}, and we will reveal the real type signature of \lstinline{*} in Section~\ref{Ad Hoc Polymorphism}.

\begin{lstlisting}[float={h t b p},caption={The hypothetical type signature of multiplication operator for \lstinline{DoubleLayer} and \lstinline{DoubleWeight}},label={hypothetical differentiable types}]
trait DoubleLayer {
  // Scalar multiplication
  def *(rhs: Double): DoubleLayer
  def *(rhs: DoubleLayer): DoubleLayer
  def *(rhs: DoubleWeight): DoubleLayer

  // Element-wise multiplication
  def *(rhs: INDArray): INDArrayLayer
  def *(rhs: INDArrayLayer): INDArrayLayer
  def *(rhs: INDArrayWeight): INDArrayLayer
}
trait DoubleWeight {
  // Scalar multiplication
  def *(rhs: Double): DoubleLayer
  def *(rhs: DoubleLayer): DoubleLayer
  def *(rhs: DoubleWeight): DoubleLayer

  // Element-wise multiplication
  def *(rhs: INDArray): INDArrayLayer
  def *(rhs: INDArrayLayer): INDArrayLayer
  def *(rhs: INDArrayWeight): INDArrayLayer
}
\end{lstlisting}

Table~\ref{differentiable types} lists DeepLearning.scala built-in differentiable types other than \lstinline{DoubleLayer} and \lstinline{DoubleWeight}.

\begin{table}[h t b p]
  \caption{Built-in Differentiable Types}\label{differentiable types}
  \begin{tabular}{c l l l}
    \toprule
    & non-trainable value & \gls{trainable variable} & \gls{differentiable expression} \\
    \midrule
    single-precision scalar & \href{https://www.scala-lang.org/api/current/scala/Double.html}{\lstinline$Double$} & \href{https://javadoc.io/page/com.thoughtworks.deeplearning/plugins-doubleweights_2.11/latest/com/thoughtworks/deeplearning/plugins/DoubleWeights%24DoubleWeight.html}{\lstinline$DoubleWeight$} & \href{https://javadoc.io/page/com.thoughtworks.deeplearning/plugins-doublelayers_2.11/latest/com/thoughtworks/deeplearning/plugins/DoubleLayers%24DoubleLayer.html}{\lstinline$DoubleLayer$} \\
    double-precision scalar & \href{https://www.scala-lang.org/api/current/scala/Float.html}{\lstinline$Float$} & \href{https://javadoc.io/page/com.thoughtworks.deeplearning/plugins-floatweights_2.11/latest/com/thoughtworks/deeplearning/plugins/FloatWeights%24FloatWeight.html}{\lstinline$FloatWeight$} & \href{https://javadoc.io/page/com.thoughtworks.deeplearning/plugins-floatlayers_2.11/latest/com/thoughtworks/deeplearning/plugins/FloatLayers%24FloatLayer.html}{\lstinline$FloatLayer$} \\
    vector & \href{https://nd4j.org/doc/org/nd4j/linalg/api/ndarray/INDArray.html}{\lstinline$INDArray$} & \href{https://javadoc.io/page/com.thoughtworks.deeplearning/plugins-indarrayweights_2.11/latest/com/thoughtworks/deeplearning/plugins/INDArrayWeights%24INDArrayWeight.html}{\lstinline$INDArrayWeight$} & \href{https://javadoc.io/page/com.thoughtworks.deeplearning/plugins-indarraylayers_2.11/latest/com/thoughtworks/deeplearning/plugins/INDArrayLayers%24INDArrayLayer.html}{\lstinline$INDArrayLayer$} \\
    \bottomrule
  \end{tabular}
\end{table}

In addition to differentiable operations, \lstinline{Layer}s and \lstinline{Weight}s can be evaluated with a \lstinline{predict} method, thus, the model can predict the next Integer by calling an ordinary function, as shown in Listing~\ref{predict}. 
You may notice the \lstinline{blockingAwait} suffix appended to \lstinline{predict}, because \lstinline{predict} returns a \lstinline{Future[Double]}
\footnote{ 
  For readers familiar to Haskell, you can understand \lstinline{Future} from the corresponding types in Haskell:
  \begin{itemize}
    \item \lstinline{Future} is an opaque type alias of a \lstinline{TryT}-transformed \lstinline{UnitContinuation}, which is used for asynchronous operations with the ability of exception handling.
    \item ``opaque type alias'' is similar to the \lstinline{newtype} feature in Haskell.
    \item \lstinline{TryT} provides the ability of exception handling, which is similar to \lstinline{ExceptT} monad transformer in Haskell.
    \item \lstinline{UnitContinuation} is similar to \lstinline{Cont ()} in Haskell, which is used for asynchronous operations, like an asynchronous version of \lstinline{IO} in Haskell.
  \end{itemize}
  All the above types are general purpose libraries, not parts of DeepLearning.scala. We use those continuation based monadic data types to archive the ability of parallel and asynchronous execution.
}
, which contains the asynchronous task to compute the result. The actual computation is not performed in  \lstinline{predict} until \lstinline{blockingAwait} is invoked\footnote{
  When \lstinline{predict} is used in a real world scenario (e.g. running a neural network in a web service), \lstinline{blockingAwait} should be replaced to \lstinline{flatMap}, instead of blocking the current thread, which is an expensive resource. We use this \lstinline{blockingAwait} here because it's more straightforward for understanding.
}.

\begin{lstlisting}[float={h t b p},caption={Inference on an untrained model},label={predict}]
val question = Seq(42.0, 43.0, 44.0)
println(guessNextNumber(question).predict.blockingAwait)
\end{lstlisting}

However, \lstinline{guessNextNumber} returns an incorrect result because the weights and bias were randomly initialized, and have not been trained.

In order to train them, a loss function is necessary:

\begin{lstlisting}[float={h t b p},caption={The differentiable square loss function},label={squareLoss}]
def squareLoss(robotAnswer: DoubleLayer, expectedAnswer: Double): DoubleLayer = {
  val difference: DoubleLayer = robotAnswer - expectedAnswer
  difference * difference
}
\end{lstlisting}

The above loss function \lstinline{squareLoss} determines the squared error between robot's answer and the correct answer.

Both \lstinline{squareLoss} and \lstinline{guessNextNumber} are ordinary functions, and can be composed in other functions:

\begin{lstlisting}[float={h t b p},caption={A differentiable function to train a linear regression model}]
def linearRegression(question: Seq[Double], expectedAnswer: Double): DoubleLayer = {
  val robotAnswer = guessNextNumber(question)
  squareLoss(robotAnswer, expectedAnswer)
}
\end{lstlisting}

\lstinline{linearRegression}, composed of \lstinline{guessNextNumber} and \lstinline{squaredLoss}, returns a \lstinline{DoubleLayer} of the loss for a specific question and its expected answer. \lstinline{linearRegression} is a linear regression model with a square loss, and it can be trained as shown Listing~\ref{train}. The \lstinline{blockingAwait} is invoked because \lstinline{train} returns a \lstinline{Future[Double]} as well.

\begin{lstlisting}[float={h t b p},caption={Training for 500 iterations},label={train}]
val question1 = Seq(3.0, 4.0, 5.0)
val expectedAnswer1 = 6.0

val question2 = Seq(13.0, 19.0, 25.0)
val expectedAnswer2 = 31.0

for (iteration <- 0 until 500) {
  linearRegression(question1, expectedAnswer1).train.blockingAwait
  linearRegression(question2, expectedAnswer2).train.blockingAwait
}
\end{lstlisting}

The weights and bias referenced by \lstinline{linearRegression} are modified during 500 iterations of training, by stochastic gradient descent, to minimize the loss returned from \lstinline{linearRegression}.

When weights and bias have been trained to make loss close to zero, \lstinline{guessNextNumber} should return values that are very close to the expected answers.

\begin{lstlisting}[float={h t b p},caption={Inference on a trained model},label={predict_trained}]
val question = Seq(42.0, 43.0, 44.0)
println(guessNextNumber(question).predict.blockingAwait)
\end{lstlisting}

This time, it will print a number clos to 45, as the model has finally learned the pattern of arithmetic progression.

The example from Listing~\ref{guessNextNumber} to Listing~\ref{predict_trained} demonstrated some basic concepts in DeepLearning.scala.

\begin{itemize}
  \item \lstinline{guessNextNumber}, \lstinline{squareLoss} and \lstinline{linearRegression} are \glspl{differentiable function} that return \glspl{differentiable expression}, which are \gls{computational graph} nodes that can be evaluated when \href{https://javadoc.io/page/com.thoughtworks.deeplearning/deeplearning_2.11/latest/com/thoughtworks/deeplearning/DeepLearning.html#train(differentiable:Differentiable)(implicitmonoid:algebra.ring.MultiplicativeMonoid[DeepLearning.this.Delta]):com.thoughtworks.future.Future[DeepLearning.this.Data]}{\lstinline{train}}ing or \href{https://javadoc.io/page/com.thoughtworks.deeplearning/deeplearning_2.11/latest/com/thoughtworks/deeplearning/DeepLearning.html#predict(differentiable:Differentiable):com.thoughtworks.future.Future[DeepLearning.this.Data]}{\lstinline{predict}}ing.
  % TODO: TRAINING/PREDICTING SHOULD BE FULLY HYPERLINKED
  \item \Glspl{differentiable expression} and \glspl{trainable variable} can be used as if they are ordinary non-differentiable values. For example, as shown in Listing~\ref{squareLoss}, you can perform scalar subtraction and multiplication between \lstinline{DoubleWeight}, \lstinline{DoubleLayer} and ordinary \lstinline{scala.Double}.
  \item When \href{https://javadoc.io/page/com.thoughtworks.deeplearning/deeplearning_2.11/latest/com/thoughtworks/deeplearning/DeepLearning.html#train(differentiable:Differentiable)(implicitmonoid:algebra.ring.MultiplicativeMonoid[DeepLearning.this.Delta]):com.thoughtworks.future.Future[DeepLearning.this.Data]}{\lstinline{train}}ing a \gls{differentiable expression}, it returns a \href{https://javadoc.io/page/com.thoughtworks.future/future_2.11/latest/com/thoughtworks/future%24%24Future.html}{\lstinline{Future}}, which encapsulates the side-effect of adjusting \glspl{trainable variable} referenced by the \gls{differentiable function}.
  \item If a \gls{differentiable function} invokes another \gls{differentiable function}, then \glspl{trainable variable} trained by one \gls{differentiable function} affect another one. For example, when training the \gls{differentiable function} \lstinline{linearRegression}, The \glspl{trainable variable} \lstinline{weights} and \lstinline{bias} are modified, hence \lstinline{guessNextNumber} automatically gains the ability to \href{https://javadoc.io/page/com.thoughtworks.deeplearning/deeplearning_2.11/latest/com/thoughtworks/deeplearning/DeepLearning.html#predict(differentiable:Differentiable):com.thoughtworks.future.Future[DeepLearning.this.Data]}{\lstinline{predict}} correct answers.
\end{itemize}

\section{Dynamic Neural Networks}

DeepLearning.scala supports dynamic neural networks. It means that the control flow of a neural network can differ according to values of internal nodes of the computational graph, when processing a specific batch of input. This is the key feature of recent deep learning frameworks like PyTorch\cite{paszke2017pytorch} or Chainer\cite{tokui2015chainer}. Especially, dynamic deep neural networks can be more efficient by skipping part of the model\cite{liu2017dynamic}.

In this section, we will present how to create a simple dynamic neural network in DeepLearning.scala, which can be considered as a simplified version of outrageously large neural networks\cite{shazeer2017outrageously}.

Suppose we have two sub-neural networks, \lstinline{leftSubnet} and \lstinline{rightSubnet} (Listing~\ref{sub-networks})\footnote{
  For performance purpose, instead of \lstinline{Seq[DoubleLayer]}, we use \lstinline{INDArrayLayer} as the type of \gls{differentiable expression} backed by ND4J. \lstinline{INDArrayLayer} supports differentiable version of most operations that ND4J's n-dimensional array \lstinline{INDArray} supports.
}. We want to build a ``gated'' network, which conditionally runs either \lstinline{leftSubnet} or \lstinline{rightSubnet} for a special \lstinline{input}.

\begin{lstlisting}[float={h t b p},caption={Predefined sub-networks},label={sub-networks}]
def leftSubnet(input: INDArrayLayer): INDArrayLayer
def rightSubnet(input: INDArrayLayer): INDArrayLayer
\end{lstlisting}

Which sub-network is selected for the \lstinline{input} should be determined by the \lstinline{gate} network, which returns a pair of differentiable double expressions that indicate the preferences between \lstinline{leftSubnet} and \lstinline{rightSubnet}, shown in (Listing~\ref{gate}).

\begin{lstlisting}[float={h t b p},caption={Predefined gate network},label={gate}]
def gate(input: INDArrayLayer): (DoubleLayer, DoubleLayer)
\end{lstlisting}

The differentiable operations on \lstinline{DoubleLayer}s  and \lstinline{INDArrayLayer}s form a differentiable embedded DSL inside the meta-language Scala. Thus, the concept of ``gated'' neural network can be considered as a conditional control flow in the differentiable DSL, and the values of nodes of a gated neural network are simply some let bindings in the DSL.

The control flow of the gated network that we want to build is described in Function~\ref{GatedNetwork}.

\begin{function}[H]
  \caption{GatedNetwork()\label{GatedNetwork}}
  \KwIn{Features extracted by preceding layers}
  \KwOut{Features passed to succeeding layers}
  \SetKwFunction{gate}{gate}
  \SetKwFunction{leftSubnet}{leftSubnet}
  \SetKwFunction{rightSubnet}{rightSubnet}
  scores \(\leftarrow\) \gate{Input}\;
  \eIf{score of left sub-network \(>\) score of right sub-network}{
      \KwRet{score of left sub-network \(\times\) \leftSubnet{Input}}\;
  }{
      \KwRet{score of right sub-network \(\times\) \rightSubnet{Input}}\;
  }
\end{function}

In DeepLearning.scala, there are three different approaches to implement the gated network. Examples of these approaches are introduced in following Section~\ref{eager}, Section~\ref{monadic}, and Section~\ref{applicative}.

\subsection{Eager Execution (bad)}
\label{eager}

An obvious approach to create the gated network is to eagerly execute the \lstinline{gate}, shown in Listing~\ref{naiveGatedNet}:

\begin{lstlisting}[float={h t b p},caption={The eager execution implementation of gated network}, label={naiveGatedNet}]
def naiveGatedNet(input: INDArrayLayer): INDArrayLayer = {
  val scores = gate(input)
  if (scores._1.predict.blockingAwait > scores._2.predict.blockingAwait) {
    scores._1 * leftSubnet(input)
  } else {
    scores._2 * rightSubnet(input)
  }
}
\end{lstlisting}

There are three sub-networks in the \lstinline{naiveGatedNet} function. The \lstinline{gate} returns a pair of \lstinline{DoubleLayer}s. By blocking await the \lstinline{predict}ion result, we get two \lstinline{Double}s, which can be used to determine which sub-network is preferred between \lstinline{leftSubnet} and \lstinline{rightSubnet}. The chosen sub-network will be multiplied with the value returned by the \lstinline{gate} in order to enabling backpropagation on the \lstinline{gate}.

However, there is a performance issue in the \lstinline{naiveGatedNet}.

In DeepLearning.scala, all differentiable expressions, including the scalar \lstinline{DoubleLayer} and vectorized \lstinline{INDArrayLayer}, contain some lazily evaluated differentiable \gls{computational graph} nodes, which will not be executed until a \lstinline{predict} or \lstinline{train} task is performed in a \lstinline{blockingAwait} or \lstinline{onComplete} call.

So, the two \lstinline{predict.blockingAwait} calls in the \lstinline{if} will execute the \gls{computational graph} in \lstinline{gate} twice. Also the \gls{computational graph} in \lstinline{naiveGatedNet} will be executed when users call \lstinline{predict} or \lstinline{train} in the future. Even worse, \lstinline{input} contains a \gls{computational graph}, too. Along with \lstinline{gate}, it will be evaluated three times, which may contain a complex future extracting process.

\subsection{Monadic Control Flow (good)}
\label{monadic}

Ideally, the calls to \lstinline{predict} should be avoided in differentiable functions. The recommended approach to create a dynamic neural network is using \href{https://javadoc.io/page/com.thoughtworks.deeplearning/deeplearning_2.11/latest/com/thoughtworks/deeplearning/DeepLearning.html#forward(differentiable:Differentiable):com.thoughtworks.raii.asynchronous.Do[com.thoughtworks.deeplearning.DeepLearning.Tape[DeepLearning.this.Data,DeepLearning.this.Delta]]}{\lstinline{forward}}, which returns a monadic value of \lstinline{Do[Tape[Data, Delta]]}, which can be used in a monadic control flow via Scalaz\cite{kenji2017scalaz}'s type classes\cite{oliveira2010type} \lstinline{Monad} and \lstinline{Applicative}.

Listing~\ref{monadicGatedNet} shows the monadic control flow of a gated network. The gated network is built from the monadic expression \lstinline{gatedForward}, which contains some \lstinline{forward} calls, which are asynchronous operations (or \href{https://javadoc.io/page/com.thoughtworks.raii/asynchronous_2.11/latest/com/thoughtworks/raii/asynchronous%24%24Do.html}{\lstinline{Do}}) that produce Wengert list records (or \href{https://javadoc.io/page/com.thoughtworks.deeplearning/deeplearning_2.11/latest/com/thoughtworks/deeplearning/DeepLearning%24%24Tape.html}{\lstinline{Tape}}s). The implementation details of \lstinline{Do} and \lstinline{Tape} will be discussed in Section~\ref{implementation}. For now, we only need to know that \lstinline{Do} is a monadic data type that supports \lstinline{flatMap}.

\begin{lstlisting}[float={h t b p},caption={Monadic gated network}, label={monadicGatedNet}]
def monadicGatedNet(input: INDArrayLayer): INDArrayLayer = {
  val scores = gate(input)
  val gatedForward: Do[Tape[INDArray, INDArray]] = {
    scores._1.forward.flatMap { tape1: Tape[Double, Double] =>
      scores._2.forward.flatMap { tape2: Tape[Double, Double] =>
        if (tape1.data > tape2.data) {
          (scores._1 * leftSubnet(input)).forward
        } else {
          (scores._2 * rightSubnet(input)).forward
        }
      }
    }
  }
  INDArrayLayer(gatedForward)
}
\end{lstlisting}

By \lstinline{flatMap}ping those \lstinline{forward} operations together, we built the entire monadic control flow \lstinline{gatedForward} for the gated network.

The \lstinline{monadicGatedNet} represents a dynamic neural network, since each \lstinline{forward} operation is started after its previous \lstinline{forward} is done. This behavior allows for dynamically determining succeeding operations according to results of previous \lstinline{forward} operations, as shown in the \lstinline{if} clause in Listing~\ref{monadicGatedNet}.

However, \lstinline{flatMap} prevents additional optimization, too.
\lstinline{scores._2.forward} have to wait for \lstinline{scores._1.forward}'s result, even if the two operations are logically independent.

\subsection{Parallel Applicative Control Flow + Sequential Monadic Control Flow (better)}
\label{applicative}

Ideally, the independent operations \lstinline{scores._1.forward} and \lstinline{scores._2.forward} should run in parallel. This can be done by tagging \lstinline{Do} as \href{https://javadoc.io/page/org.scalaz/scalaz_2.11/latest/scalaz/Tags%24%24Parallel.html}{\lstinline{Parallel}}, and use \href{https://javadoc.io/page/org.scalaz/scalaz_2.11/latest/scalaz/Applicative.html#tuple2[A,B](fa:=>F[A],fb:=>F[B]):F[(A,B)]}{\lstinline{scalaz.Applicative.tuple2}}\cite{mcbride2008applicative} instead of \lstinline{flatMap} (Listing~\ref{applicativeMonadicGatedNet}).

\begin{lstlisting}[float={h t b p},caption={Applicative + monadic gated network}, label={applicativeMonadicGatedNet}]
def applicativeMonadicGatedNet(input: INDArrayLayer): INDArrayLayer = {
  val scores = gate(input)
  val parallelForward1: ParallelDo[Tape[Double, Double]] = {
    Parallel(scores._1.forward)
  }
  val parallelForward2: ParallelDo[Tape[Double, Double]] = {
    Parallel(scores._2.forward)
  }
  val Parallel(stage1) = {
  	parallelForward1.tuple2(parallelForward2)
  }
  def stage2(tapes: (Tape[Double, Double], Tape[Double, Double])) = {
    if (tapes._1.data > tapes._2.data) {
      (scores._1 * leftSubnet(input)).forward
    } else {
      (scores._2 * rightSubnet(input)).forward
    }
  }

  val gatedForward = stage1.flatMap(stage2)
  INDArrayLayer(gatedForward)
}
\end{lstlisting}

This \lstinline{applicativeMonadicGatedNet} takes both advantages from applicative functors and monads. The entire control flow is a \lstinline{flatMap} sequentially composed of two stages. In \lstinline{stage1}, there is a \lstinline{tuple2} composed of \lstinline{scores._1.forward} and \lstinline{scores._2.forward} in parallel. Then, in \lstinline{stage2}, the succeeding operation is dynamically determined according to \lstinline{tapes}, the result of \lstinline{stage1}.

The parallel applicative operation is also the default behavior for all built-in vector binary operators. Listing~\ref{parallelByDefault} shows some simple expressions that will be executed in parallel.

\begin{lstlisting}[float={h t b p},caption={By default, \lstinline{a * b} and \lstinline{c * d} will be executed in parallel because they are independent}, label={parallelByDefault}]
def parallelByDefault(a: INDArrayLayer, b: INDArrayLayer, c: INDArrayLayer, d: INDArrayLayer): INDArrayLayer = {
  a * b + c * d
}
\end{lstlisting}

By combining both applicative functors and monads, DeepLearning.scala supports dynamic neural network and still allows the independent parts of the neural network to run in parallel. In addition, the backward pass of differentiable functions built from parallel applicative functors or built-in vector binary operators will be executed in parallel, too.

\subsection{Direct style DSL for Applicative and Monadic Control Flow (best)}

In the previous section, we had presented a dynamic neural network executed in parallel. However, the usage of \lstinline{flatMap} and \lstinline{Parallel}-tagged types may scare algorithm authors who are not familiar with monadic programming. Ideally, the code written by those people should look straightforward and has the same structure in pseudo-code~\ref{GatedNetwork} or Listing~\ref{naiveGatedNet}, and still gain the benefits of Listing~\ref{applicativeMonadicGatedNet}.

The goal can be achieved by transforming the direct style code into monadic and applicative code at compile-time. We created an DSL with the help of the !-notation provided by Dsl.scala\cite{yang2017dsl}, which provides Scala compiler plugins to perform the necessary compiler-time transformation.

As shown in Listing~\ref{dslGatedNet}, the !-notation ``extracts'' the \lstinline{Double} values from a pair of \lstinline{DoubleLayer}s in parallel. Those \lstinline{Double}s are ordinary non-differentiable Scala types that will not backpropagate, and can be used in ordinary Scala control flow expression like \lstinline{if}.

\begin{lstlisting}[float={h t b p},caption={Dsl.scala powered direct style gated network}, label={dslGatedNet}]
def dslGatedNet(input: INDArrayLayer): INDArrayLayer = {
  val scoreLayers: (DoubleLayer, DoubleLayer) = gate(input)
  val scores: (Double, Double) = !scoreLayers
  if (scores._1 > scores._2) {
    scoreLayers._1 * leftSubnet(input)
  } else {
    scoreLayers._2 * rightSubnet(input)
  }
}
\end{lstlisting}

Generally, a !-notation on a \lstinline{Layer} will generate a monadic \lstinline{flatMap} call, to extract the value of forward pass of the \lstinline{Layer}; a !-notation on a tuple of \lstinline{Layer}s will generate some applicative \lstinline{<*>} and \lstinline{map} calls, to extract a tuple of values of forward pass of those \lstinline{Layer}s, in parallel. Thus, the actually code generated by Dsl.scala's compiler plugins for Listing~\ref{dslGatedNet} is similar to Listing~\ref{applicativeMonadicGatedNet}.

\section{Ad Hoc Polymorphic Differentiable Functions}
\label{Ad Hoc Polymorphism} 

In section~\ref{concepts}, we had presented the hypothetical differentiable types of multiplication for \lstinline{DoubleLayer} and \lstinline{DoubleWeight}. However, the method overloading approach shown in Listing~\ref{hypothetical differentiable types} is too verbose, and requires a lot of boilerplate code. In this section, we will present an approach to create custom \glspl{differentiable function}, without such methods overloading.

A \gls{differentiable function} is a neural network. 
Ideally, a differentiable function should be an ad hoc polymorphic function that accepts heterogeneous types of parameters, including:

\begin{itemize}
  
  \item A vanilla vector input. i.e. \lstinline{INDArray}.

  \item \Glspl{differentiable expression} of hidden states produced by any previous neural network layers, i.e. any \lstinline{INDArrayLayer}s regardless of the prefixes.
  
  \item \Glspl{trainable variable} in the case of activation maximization technique\cite{erhan2009visualizing}.i.e. any \lstinline{INDArrayWeight}s regardless of the prefixes.
  
  \item Other user defined differentiable type.

\end{itemize}

Table~\ref{differentiable types} shows nine types that have built-in \href{https://javadoc.io/page/com.thoughtworks.deeplearning/deeplearning_2.11/latest/com/thoughtworks/deeplearning/DeepLearning.html}{\lstinline{DeepLearning}} type classes.

This can be achieved with the help of\cite{gurnelltype}'s Scala encoding of dependent-type type class. We defined a \href{https://javadoc.io/page/com.thoughtworks.deeplearning/deeplearning_2.11/latest/com/thoughtworks/deeplearning/DeepLearning.html}{\lstinline{DeepLearning}} (Listing~\ref{DeepLearning}\footnote{
For readers who are more familiar with Idris, there is a corresponding notation in Idris:
\par\usebox{\IdrisDeepLearning}
}) type class that witnesses any supported expressions including \glspl{differentiable expression}, \glspl{trainable variable}, or vanilla non-differentiable types. The users can create type aliases to restrict the types of state during forward pass and backward pass as shown in Listing~\ref{INDArrayDeepLearning}.

\begin{lstlisting}[float={h t b p},caption={The dependent-type type \lstinline{DeepLearning}}, label={DeepLearning}]
trait DeepLearning[Differentiable] {
  /** The type of result calculated during forward pass */
  type Data

  /** The type of derivative during backward pass */
  type Delta

  def forward(differentiable: Differentiable): Do[Tape[Data, Delta]]
  // Other auxiliary methods is omitted
}
\end{lstlisting}

\begin{lstlisting}[float={h t b p},caption={A type class alias that witnesses dense vector expressions}, label={INDArrayDeepLearning}]
type INDArrayExpression[Expression] = DeepLearning[Expression] {
  /** The type of result calculated during forward pass */
  type Data = INDArray

  /** The type of derivative during backward pass */
  type Delta = INDArray
}
\end{lstlisting}

By using \lstinline{INDArrayExpression} as a context bound, we can create a polymorphic differentiable function that accepts any vector expression.

\begin{lstlisting}[float={h t b p},caption={A polymorphic differentiable function}, label={polymorphicDifferentiableFunction}]
def polymorphicDifferentiableFunction[A: INDArrayExpression, B: INDArrayExpression, C: INDArrayExpression, D: INDArrayExpression](a: A, b: B, c: C, d: D): INDArrayLayer = {
  a * b + c * d
}
\end{lstlisting}

Listing~\ref{polymorphicDifferentiableFunction} is similar to Listing~\ref{parallelByDefault}, except each argument of \lstinline{polymorphicDifferentiableFunction} accepts \lstinline{INDArray}, \lstinline{INDArrayWeight} or \lstinline{INDArrayLayer} respectively, not only  \lstinline{INDArrayLayer}.

Built-in operations including arithmetic operations, \lstinline{max}, and \lstinline{dot} are polymorphic differentiable functions defined in this approach, too, which can be used in user-defined polymorphic \glspl{differentiable function}.

\section{Implementation}
\label{implementation}

In this section, we will introduce the internal data structure used in DeepLearning.scala to perform AD.

\begin{itemize}

  \item For ease of understanding, Section~\ref{dual number} starts from a simple dual number implementation \lstinline{DualNumber}, which was known as an approach to perform forward mode AD for scalar values.
  
  \item Section~\ref{monadic dual number} introduces our variation of dual number \lstinline{ClosureBasedDualNumber}, which supports tree-structured reverse mode AD (aka backpropagation) for multiple \glspl{trainable variable}.

  \item Section~\ref{generic tape} shows the actual data type \lstinline{Tape} in DeepLearning.scala, which is generalized to not only scalar types, but also vector types and any other differentiable types.

  \item Section~\ref{reference counted tape} discovers the monadic control flow \lstinline{Do}, which manages the life cycle of \lstinline{Tape}s, sharing \lstinline{Tape}s for common \gls{computational graph} nodes, allowing for an arbitrary DAG(Directed Acyclic Graph)-structured \gls{computational graph}.

  \item Section~\ref{training iteration} summarizes the entire execution process during a training iteration, showing how the user-defined \glspl{differentiable function} get executed through internal mechanisms \lstinline{Do} and \lstinline{Tape}.

\end{itemize}

\subsection{Ordinary Dual Number}
\label{dual number}

Our approach for reverse mode AD uses a data structure similar to traditional forward mode AD, with only a few changes.

Forward mode AD can be considered as computation on dual number. For example, dual number for scalar types can be implemented as Listing~\ref{DualNumber}:

\begin{lstlisting}[float={h t b p},caption={Dual number for forward mode AD}, label={DualNumber}]
type Data = Double
type PartialDelta = Double
case class DualNumber(data: Data, delta: PartialDelta)
\end{lstlisting}

Arithmetic operations on those dual number can be implemented as Listing~\ref{dualArithmetic}:

\begin{lstlisting}[float={h t b p},caption={Arithmetic operations on dual number}, label={dualArithmetic}]
object DualNumber {
  def plus(left: DualNumber, right: DualNumber): DualNumber = {
    DualNumber(left.data + right.data, left.delta + right.delta)
  }
  def multiply(left: DualNumber, right: DualNumber): DualNumber = {
    DualNumber(left.data * right.data, left.data * right.delta + right.data * left.delta)
  }
}
\end{lstlisting}

\subsection{Monadic Closure-based Dual Number}
\label{monadic dual number}

However, it is hard to type this approach if we want to support multiple \glspl{trainable variable}, with the number unknown before runtime. \lstinline{PartialDelta} in Listing~\ref{DualNumber} represents the partial derivative of \glspl{trainable variable}. In AD tools that support only one \gls{trainable variable}, the \gls{trainable variable} is usually forced to be the input. Hence \lstinline{PartialDelta} is the input type for those AD tools. This assumption is broken for our case, since our delta type of a specific \lstinline{DualNumber} must contain derivatives for all \glspl{trainable variable} that were used to produce the \lstinline{DualNumber}, not only the partial derivative of input. As a result, the type of delta varies when the number of \glspl{trainable variable} grows.

To type-check the delta, considering the only usage of the delta in a neural network is in updating \glspl{trainable variable} in a gradient descent based optimization algorithm, we can replace \lstinline{PartialDelta} to an \lstinline{UpdateWeights} closure as shown in Listing~\ref{class ClosureBasedDualNumber}.

\begin{lstlisting}[float={h t b p},caption={Replacing \lstinline{PartialDelta} to a closure}, label={class ClosureBasedDualNumber}]
type Data = Double  
case class ClosureBasedDualNumber(data: Data, backward: UpdateWeights)
\end{lstlisting}

Unlike \lstinline{PartialDelta}, \lstinline{UpdateWeights} is not a number that supports native arithmetic operations, thus we have to replace native arithmetic operations on \lstinline{PartialDelta} to some static functions on \lstinline{UpdateWeights} when implement arithmetic operations for the new dual number, as shown in Listing~\ref{object ClosureBasedDualNumber}

\begin{lstlisting}[float={h t b p},caption={Replacing operations on \lstinline{PartialDelta} to custom functions for \lstinline{UpdateWeights}}, label={object ClosureBasedDualNumber}]  

object ClosureBasedDualNumber {
  def plus(left: ClosureBasedDualNumber, right: ClosureBasedDualNumber): ClosureBasedDualNumber = {
    ClosureBasedDualNumber(left.data + right.data, UpdateWeights.plus(left.backward(), right.backward()))
  }
  def multiply(left: ClosureBasedDualNumber, right: ClosureBasedDualNumber): ClosureBasedDualNumber = {
    ClosureBasedDualNumber(
      left.data * right.data,
      UpdateWeights.multiply(left.data, right.backward()) + UpdateWeights.multiply(right.data, left.backward()))
  }
}
\end{lstlisting}

The only question remaining is to implement the \lstinline{UpdateWeights}, to make its behavior be equivalent to the original dual number implementation.

Mathematically, the \lstinline{UpdateWeights} type in a dual number should form any vector space, i.e. the \lstinline{UpdateWeights} closure itself must support addition and scalar multiplication operations.

Our approach is making \lstinline{UpdateWeights} be a function type that contains side-effects to update \glspl{trainable variable}.  Thus the addition operation for closures can be defined as (\ref{differentiable addition}).

\begin{equation}
\label{differentiable addition}
(f_0 + f_1)(x) = f_0(x) + f_1(x)
\end{equation}

And the scalar multiplication operation for closures is defined as (\ref{differentiable multiplication}):

\begin{equation}
\label{differentiable multiplication}
(x_0f)(x_1) = f(x_0x_1)
\end{equation}

The above definition of arithmetic operations can be implemented in monadic data types as shown in Listing~\ref{UpdateWeights}.

\begin{lstlisting}[float={h t b p},caption={Arithmetic operations for the closure that contains side-effects}, label={UpdateWeights}, escapeinside={(*}{*)}]
type UpdateWeights = Do[Double] => SideEffects
object UpdateWeights {
  /**(* $(f_0 + f_1)(x) = f_0(x) + f_1(x)$ *)*/
  def plus(f0: UpdateWeights, f1: UpdateWeights) = { doX: Do[Double] =>
    f0(doX) |+| f1(doX)
  }

  /**(* $(x_0f)(x_1) = f(x_0x_1)$ *)*/
  def multiply(x0: Double, f: UpdateWeights) = { doX1: Do[Double] =>
    f(doX1.map(x0 * _))
  }
}
\end{lstlisting}

\lstinline{UpdateWeights}, as a replacement to original \lstinline{PartialDelta}, is a closure able to update derivatives for all weights with a coefficient (the \lstinline{Double} parameter).
\lstinline{|+|} is the \lstinline{append} operation of \lstinline{Semigroup}, which could be any cumulative data type.

Also note that the parameter is a monadic data type \href{https://javadoc.io/page/com.thoughtworks.raii/asynchronous_2.11/latest/com/thoughtworks/raii/asynchronous%24%24Do.html}{\lstinline{Do}} that encapsulates the computation of derivative. Unlike strictly evaluated values, \lstinline{Do} is an operation only evaluated when needed.

In DeepLearning.scala, our \lstinline{SideEffects} is based on the asynchronous operation type \texttt{UnitContinuation}. Our built-in differentiable operations execute the independent parts of backward pass in parallel with the help of \lstinline{Applicative} type class instances of \texttt{UnitContinuation}.

\begin{lstlisting}[float={h t b p},caption={Monadic side-effects}, label={SideEffects}]
type SideEffects = UnitContinuation[Unit]
\end{lstlisting}

\lstinline{UnitContinuation[A]} is an opaque alias\cite{erik2017opaque} of \lstinline{(A => Trampoline[Unit]) => Trampoline[Unit]}, implemented in a separate library at \href{https://github.com/ThoughtWorksInc/future.scala}{future.scala}. It is used in DeepLearning.scala as a monadic data type for encapsulating side effects in stack-safe asynchronous programming.

The \lstinline{SideEffects} for neural networks conform to the associative law because the only side effects is updating \glspl{trainable variable}. Thus, our \lstinline{UpdateWeights.plus} and \lstinline{UpdateWeights.multiply} are equivalent to the operations on strictly evaluated scalar value \lstinline{PartialDelta} in forward mode AD.

Since \lstinline{UpdateWeights} is a closure with side effects, a \gls{trainable variable} can be represented as a tuple of a mutable value and the action to modify the mutable value.

\begin{lstlisting}[float={h t b p},caption={Create a dual number for a \gls{trainable variable}}, label={createTrainableVariable}]
def createTrainableVariable(initialValue: Double, learningRate: Double): ClosureBasedDualNumber = {
  var data = initialValue
  val backward: UpdateWeights = { doDelta: Do[Double] =>
    val sideEffects: Do[Unit] = doDelta.map { delta =>
      data -= learningRate * delta
    }
    convertDoToContinuation(sideEffects)
  }
  ClosureBasedDualNumber(data, backward)
}
\end{lstlisting}

In Listing~\ref{createTrainableVariable}, the \gls{trainable variable} is trained by a fixed learning rate to simplify the hyperparameters of optimization algorithms. The \lstinline{sideEffects} is the only code in the entire code base that performs native side effects. The actual DeepLearning.scala implementation has a more sophisticated mechanism to customize the implementation of side effects, allowing other optimizers and custom hyperparameters.

Similar to \glspl{trainable variable}, a non-trainable value can be represented as a tuple of the value and a no-op closure as shown in Listing~\ref{createLiteral}.

\begin{lstlisting}[float={h t b p},caption={Create a dual number for a  non-trainable value}, label={createLiteral}]
def createLiteral(data: Double): ClosureBasedDualNumber = {
  val backward = { doDelta: Do[Double] =>
    UnitContinuation.now(())
  }
  ClosureBasedDualNumber(data, backward)
}
\end{lstlisting}

Because \lstinline{delta} is an action instead of pre-evaluated value, the implementation of \lstinline{backward} for a non-trainable value can entirely avoid executing any unnecessary computation in \lstinline{doDelta}.

Finally, we can create a differentiable function as shown in Listing~\ref{computationalTree}, whose leaf nodes are \lstinline{createTrainableVariable} and \lstinline{createLiteral}, and internal nodes are arithmetic operations in Listing~\ref{object ClosureBasedDualNumber}.

\begin{lstlisting}[float={h t b p},caption={A tree-structured \gls{differentiable function}},label={computationalTree}]
val w0 = createTrainableVariable(math.random, 0.001)
val w1 = createTrainableVariable(math.random, 0.001)

def computationalTree(x: ClosureBasedDualNumber) = {
  val y0 = ClosureBasedDualNumber.multiply(x, w0)
  val y1 = ClosureBasedDualNumber.multiply(y0, w1)
  y1
}
\end{lstlisting}

\begin{figure}[h t b p]
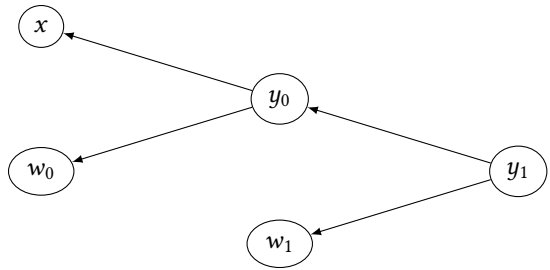


  \begin{dot2tex}[dot,mathmode]
  digraph {
    rankdir=RL
    shape=circle

    y_0 -> x
    y_0 -> w_0
    y_1 -> y_0
    y_1 -> w_1
    
  }
  \end{dot2tex}
    
  \caption{A tree-structured \gls{computational graph}}
  \label{tree}
\end{figure}

The \gls{computational graph} of \lstinline{computationalTree} is shown in Figure~\ref{tree}. Note that the arrow direction denotes the dependency between expressions, from arrow tail to arrow head, which is the reverse of the direction of data flow.

The closure-based dual number \lstinline{y1} has a closure \lstinline{backward}, which returns \lstinline{SideEffects} that recursively change all \glspl{trainable variable} referenced by the closure.

\begin{figure}[h t b p]
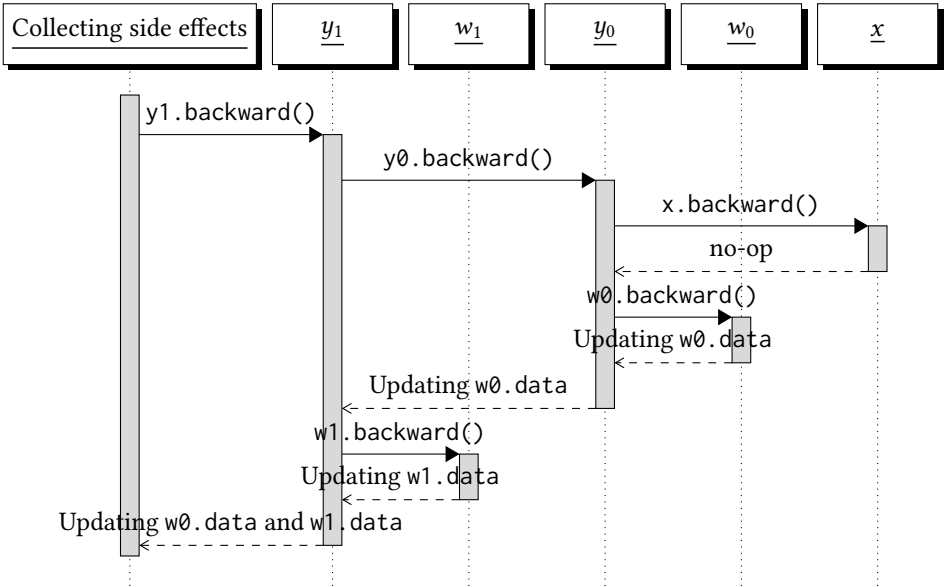

  \newcommand{\x}{$x$}
  \newcommand{\y}[1]{$y_#1$}
  \newcommand{\w}[1]{$w_#1$}
  
  \begin{sequencediagram}
    \newthread{Collecting side effects}{Collecting side effects}
    \newinst{y1}{\y1}
    \newinst{w1}{\w1}
    \newinst{y0}{\y0}
    \newinst{w0}{\w0}
    \newinst{x}{\x}
    \begin{call}{Collecting side effects}{\lstinline{y1.backward()}}{y1}{Updating \lstinline{w0.data} and \lstinline{w1.data}}
        \begin{call}{y1}{\lstinline{y0.backward()}}{y0}{Updating \lstinline{w0.data}}
          \begin{call}{y0}{\lstinline{x.backward()}}{x}{no-op}
          \end{call}
          \begin{call}{y0}{\lstinline{w0.backward()}}{w0}{Updating \lstinline{w0.data}}
          \end{call}
        \end{call}
        \begin{call}{y1}{\lstinline{w1.backward()}}{w1}{Updating \lstinline{w1.data}}
        \end{call}
    \end{call}
  \end{sequencediagram}

  \caption{Backpropagation for a tree-structured \gls{computational graph}}
  \label{tree backpropagation}
\end{figure}

Note that \lstinline{backward} itself does not perform any side effects. It just collects all side effects into a \lstinline{UnitContinuation[Unit]}. Figure~\ref{tree backpropagation} shows how the side effects of updating \glspl{trainable variable} are collected.

Finally, the collected side effects of \lstinline{UnitContinuation[Unit]} returned from \lstinline{y1.backward} can be performed by a \href{https://javadoc.io/page/com.thoughtworks.future/future_2.11/latest/com/thoughtworks/continuation%24%24UnitContinuationOps.html#blockingAwait():A}{\lstinline{blockingAwait}} or \href{https://javadoc.io/page/com.thoughtworks.future/future_2.11/latest/com/thoughtworks/continuation%24%24ContinuationOps.html#onComplete(continue:A=>R):R}{\lstinline{onComplete}} call.

\subsection{Generic Tape}
\label{generic tape}

This closured-based monadic dual number can be generalized to any linear spaces, not only scalar types such as \lstinline{Double}, but also n-dimensional arrays.

The dual number type that we actually defined in DeepLearning.scala is \lstinline{Tape}, a generic version of \lstinline{ClosureBasedDualNumber} in Listing~\ref{class ClosureBasedDualNumber}. We replaced \lstinline{ClosureBasedDualNumber}'s hard-coded \lstinline{Double} to type parameters \lstinline{Data} and \lstinline{Delta}, as shown in Listing~\ref{Tape}.

\begin{lstlisting}[float={h t b p},caption={Generic closured-based monadic dual number}, label={Tape}]
final case class Tape[+Data, -Delta](
  data: Data,
  backward: Do[Delta] => UnitContinuation[Unit]
)
\end{lstlisting}

\lstinline{Data} and \lstinline{Delta} are usually the same, but they can also be different types. For example, you can create a type whose \lstinline{Data} is a dense n-dimensional array and whose {Delta} is a pair of index and scalar, representing a dense tensor that sparsely updates.

This data structure is similar to a Wengert list in traditional reverse mode AD, except our tape is a tree of closures instead of a list.

\subsection{Reference Counted Tape}
\label{reference counted tape}

Although the closured-based dual number approach from Listing~\ref{class ClosureBasedDualNumber} to Listing~\ref{Tape} supports multiple \glspl{trainable variable}, the closure-based computation has a performance issue in the case of diamond dependencies.

Listing~\ref{y2} shows a \gls{differentiable function} \lstinline{diamondDependentComputationalGraph} that contains diamond dependencies to some \glspl{differentiable expression} or \glspl{trainable variable}.

\begin{lstlisting}[float={h t b p},caption={A diamond dependent \gls{differentiable function}}, label={y2}]
val w = createTrainableVariable(math.random, 0.001)
def diamondDependentComputationalGraph(x: ClosureBasedDualNumber) = {
  val y0 = ClosureBasedDualNumber.multiply(x, w)
  val y1 = ClosureBasedDualNumber.multiply(y0, y0)1
  val y2 = ClosureBasedDualNumber.multiply(y1, y1)
  y2
}
\end{lstlisting}

\begin{figure}[h t b p]
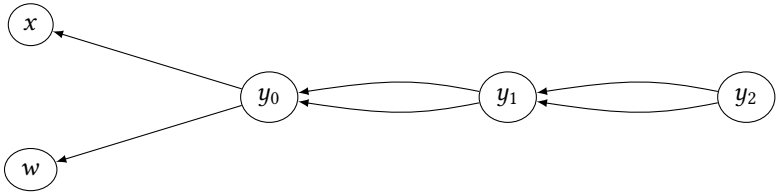


  \begin{dot2tex}[dot,mathmode]
  digraph {
    rankdir=RL
    shape=circle

	y_0->x
    y_0->w
    y_1->y_0
    y_1->y_0
    y_2->y_1
    y_2->y_1
  }
  \end{dot2tex}

  \caption{A diamond dependent \gls{computational graph}}
  \label{diamond}
\end{figure}

The \gls{computational graph} of \lstinline{diamondDependentComputationalGraph} are shown in Figure~\ref{diamond}.

When \lstinline{y2.backward} is invoked, in order to collect side effects of \lstinline{y2}'s dependencies, \lstinline{y1.backward} will be invoked, twice, and each \lstinline{y1.backward} call will triggers two \lstinline{y0.backward} calls. As a result, for each iteration of backpropagation, \lstinline{y0.backward}, \lstinline{w.backward} and \lstinline{x.backward} are invoked four times, respectively.

\begin{figure}[h t b p]
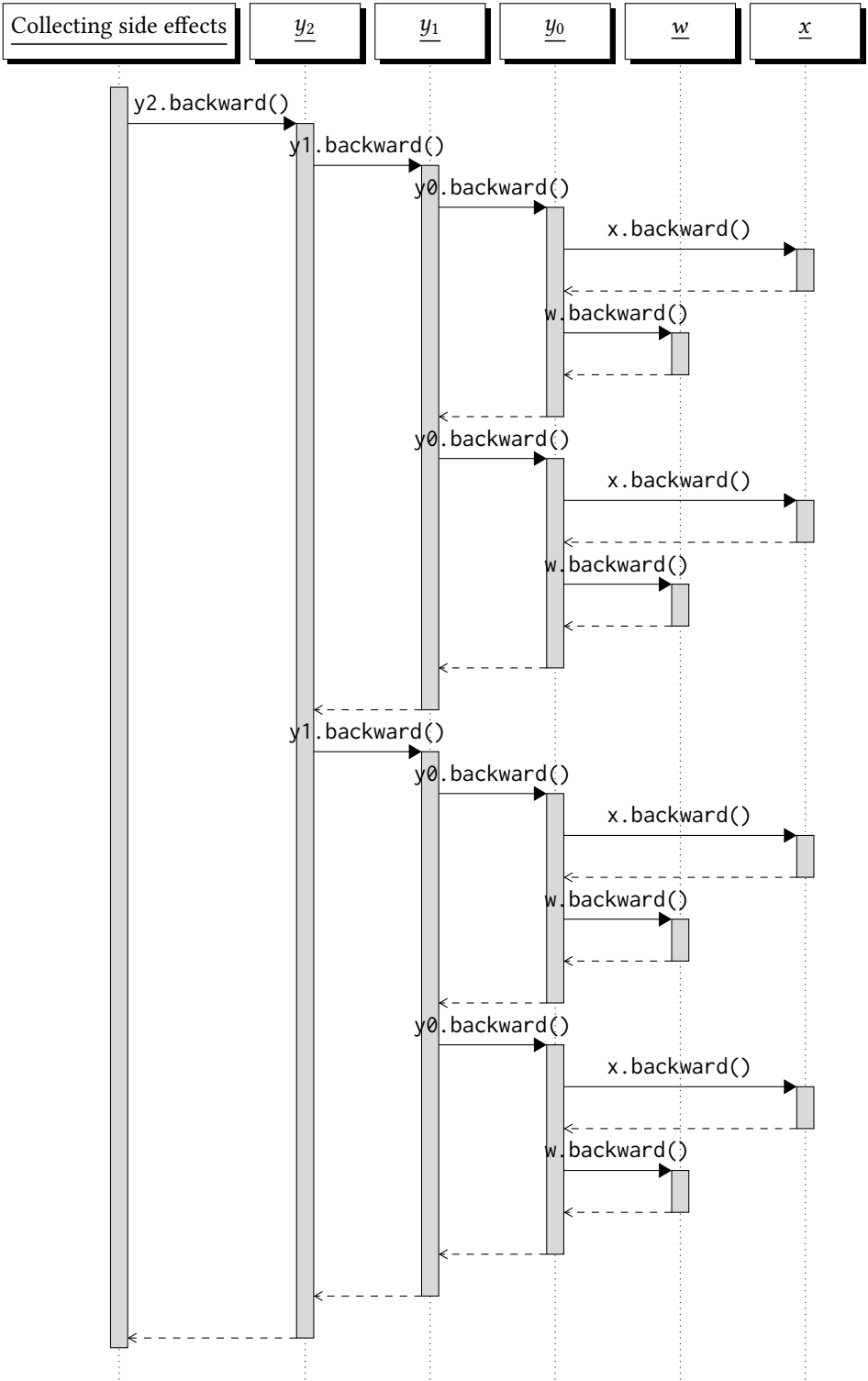

  \newcommand{\x}{$x$}
  \newcommand{\w}{$w$}
  \newcommand{\y}[1]{$y_#1$}
  
  \begin{sequencediagram}
    \newthread{Collecting side effects}{Collecting side effects}
    \newinst{y2}{\y2}
    \newinst{y1}{\y1}
    \newinst{y0}{\y0}
    \newinst{w}{\w}
    \newinst{x}{\x}

    \newcommand{\callyzero}{
      \begin{call}{y1}{\lstinline{y0.backward()}}{y0}{}
        \begin{call}{y0}{\lstinline{x.backward()}}{x}{}
        \end{call}
        \begin{call}{y0}{\lstinline{w.backward()}}{w}{}
        \end{call}
      \end{call}
    }
    
    \newcommand{\callyone}{
      \begin{call}{y2}{\lstinline{y1.backward()}}{y1}{}
        \callyzero
        \callyzero
      \end{call}
    }

    \begin{call}{Collecting side effects}{\lstinline{y2.backward()}}{y2}{}
      \callyone
      \callyone
    \end{call}
  \end{sequencediagram}

  \caption{Backpropagation for a diamond dependent \gls{computational graph}}
  \label{diamond backpropagation}
\end{figure}

The process in \lstinline{y2.backward} is shown in Figure~\ref{diamond backpropagation}.

Generally, given $n$ levels of nested diamond dependencies, the computational complexity is $O(2^n)$, which is unacceptable for neural networks that may share common \glspl{differentiable expression}.

We introduced a reference counting algorithm for dual numbers, to avoid the exponential time complexity, by only calling backward once.

The reference counting is managed in a wrapper of \lstinline{Tape}, which has additional \lstinline{acquire} and \lstinline{release} functions.

Each wrapper has two internal states:
\begin{enumerate*}
  \item reference counter,
  \item accumulator of delta.
\end{enumerate*}
Respectively, \lstinline{acquire} and \lstinline{release} calls will increase and decrease the reference counter, and \lstinline{backward} calls will cumulate the \lstinline{delta} to the accumulator.

\begin{figure}[h t b p]
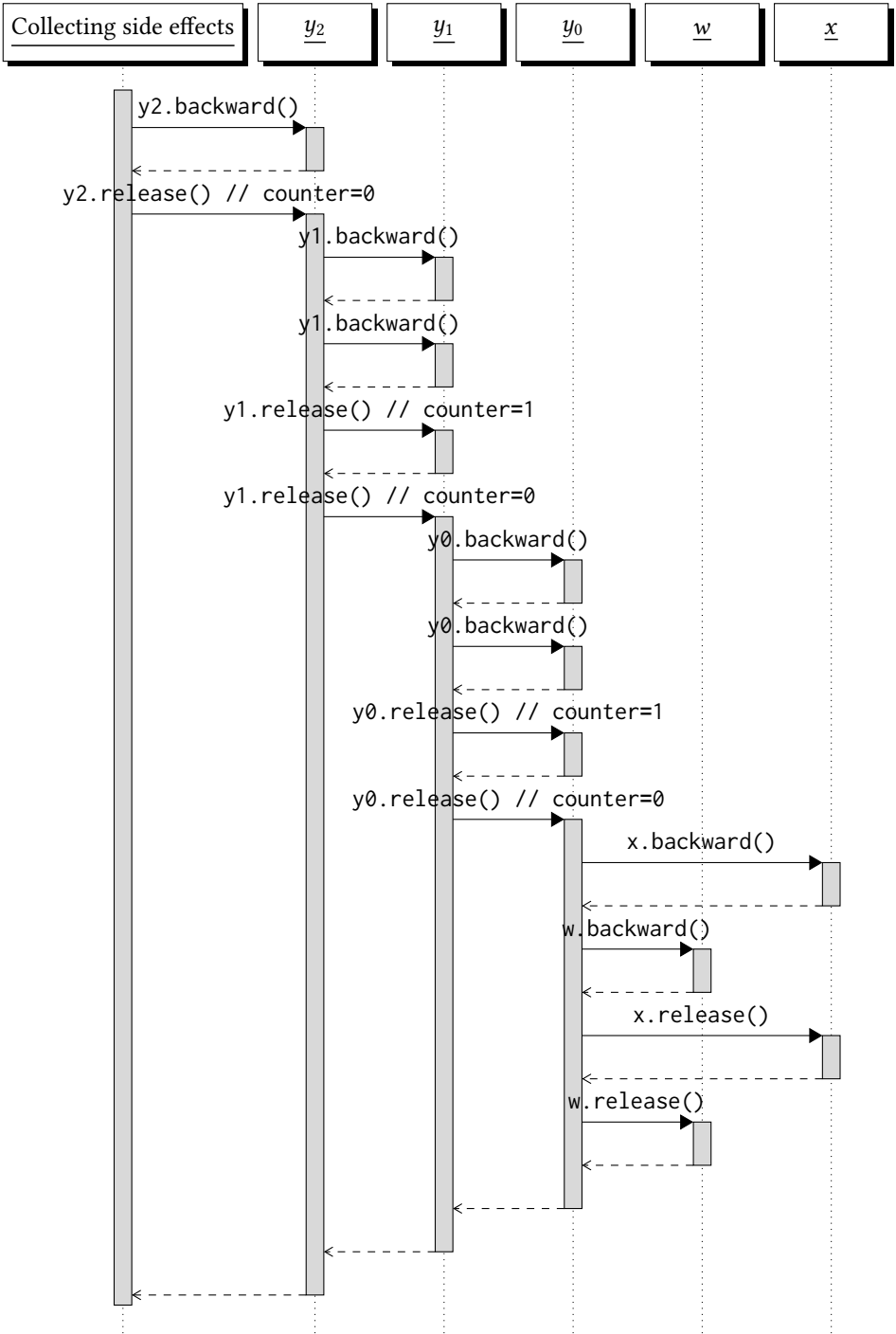

  \newcommand{\x}{$x$}
  \newcommand{\w}{$w$}
  \newcommand{\y}[1]{$y_#1$}
  
  \begin{sequencediagram}    
    \newthread{Collecting side effects}{Collecting side effects}
    \newinst{y2}{\y2}{}
    \newinst{y1}{\y1}{}
    \newinst{y0}{\y0}{}
    \newinst{w}{\w}{}
    \newinst{x}{\x}{}

    \begin{call}{Collecting side effects}{\lstinline{y2.backward()}}{y2}{}
    \end{call}
    \begin{call}{Collecting side effects}{\lstinline{y2.release() // counter=0}}{y2}{}
      \begin{call}{y2}{\lstinline{y1.backward()}}{y1}{}
      \end{call}
      \begin{call}{y2}{\lstinline{y1.backward()}}{y1}{}
      \end{call}
      \begin{call}{y2}{\lstinline{y1.release() // counter=1}}{y1}{}
      \end{call}
      \begin{call}{y2}{\lstinline{y1.release() // counter=0}}{y1}{}
        \begin{call}{y1}{\lstinline{y0.backward()}}{y0}{}
        \end{call}
        \begin{call}{y1}{\lstinline{y0.backward()}}{y0}{}
        \end{call}
        \begin{call}{y1}{\lstinline{y0.release() // counter=1}}{y0}{}
        \end{call}
        \begin{call}{y1}{\lstinline{y0.release() // counter=0}}{y0}{}
          \begin{call}{y0}{\lstinline{x.backward()}}{x}{}
          \end{call}
          \begin{call}{y0}{\lstinline{w.backward()}}{w}{}
          \end{call}
          \begin{call}{y0}{\lstinline{x.release()}}{x}{}
          \end{call}
          \begin{call}{y0}{\lstinline{w.release()}}{w}{}
          \end{call}
        \end{call}
      \end{call}
    \end{call}
  \end{sequencediagram}

  \caption{Backpropagation for a diamond dependent \gls{computational graph} (with reference counting)}\label{reference-counted backpropagation}
\end{figure}

When the reference counting algorithm is enabled, \lstinline{backward} is not recursive any more. Instead, a wrapper only call \lstinline{backward} of its dependencies when the reference counter is decreased to zero. The entire process of backpropagation is shown in Figure~\ref{reference-counted backpropagation}.

This wrapper is implemented as the monadic data type \href{https://javadoc.io/page/com.thoughtworks.raii/asynchronous_2.11/latest/com/thoughtworks/raii/asynchronous%24%24Do.html}{\lstinline{Do}}, in which the side effects of updating counters and accumulators are monadic control flows. With the help of \lstinline{Do}, now our \glspl{computational graph} are modeled in \lstinline{Do[Tape[Data, Delta]]}, which can be created by \lstinline{forward} methods described in Section~\ref{monadic}. As mentioned in Section~\ref{applicative}, \gls{computational graph} node of binary operations are evaluated in parallel.

In a traditional backpropagation implementation, tape is a list, hence both the execution order of backward pass and forward pass must be sequentially reverse to each other. Even a previous attempt of closure-based tape\cite{pearlmutter2008reverse} still requires conversion to sequential expressions of A-normal form\cite{sabry1993reasoning}. 

By introducing reference counting, the execution order of our backward pass and forward pass do not have to be exactly reverse, hence the conversion to A-normal form becomes unnecessary. As a result, DeepLearning.scala supports out-of-order execution in both forward pass and backward pass, in which the independent sub-graph can be even executed in parallel.

\subsection{The Overview of a Training Iteration}\label{training iteration}

In brief, in each iteration, a \gls{differentiable function} that contains multiple \glspl{trainable variable} can be trained in the following steps:

\begin{enumerate}
  \item Executing the user-defined \gls{differentiable function} with a batch of input, to obey a \gls{differentiable expression} (i.e. a subtype of \lstinline{Layer}).
  \label{obey}
  
  \item Calling \lstinline{forward} on \gls{differentiable expression} to build a \gls{computational graph} (i.e. a \lstinline{Do[Tape[Data, Delta]]}). The reference counter to the \gls{computational graph} is zero at the point.
  \label{build}
  
  \item Performing the \lstinline{forward} pass of \gls{differentiable expression} to build a tape (i.e. a \lstinline{Tape[Data, Delta]}), which contains a pair of the result of forward pass and a \lstinline{backward} closure. The reference counter of each node in a \gls{computational graph} is increased during this step.
  \label{acquire}

  \item Performing \lstinline{backward} closure of the root node of the \gls{computational graph}. The accumulator of delta on the root node is updated.

  \item Releasing of the root node of the \gls{computational graph}. The reference counter of each node in a \gls{computational graph} is decreased to zero and the \lstinline{backward} closure of each node is performed during this step, thus all referenced \glspl{trainable variable} are updated.
  \label{release}

\end{enumerate}

Note that step~\ref{obey} and step~\ref{build} are pure function calls, with no side effects. Step~\ref{acquire} to step~\ref{release} are monadic control flows, which encapsulate some side effects that will be performed only when an unsafe method \lstinline{blockingAwait} or \lstinline{onComplete} is eventually called.

\section{Evaluation}

We created some benchmarks to evaluate the computational performance of DeepLearning.scala, especially, we want to measure:

\begin{enumerate}
  \item How parallel execution affect the time cost of a training iteration of different structured neural networks;
  \item the computational performance impact when a part of a neural network is disabled.
\end{enumerate}

Those benchmarks are built with Jmh\cite{shipilevjmh}, running on CPU, measuring the number of mini-batches per second, for training neural networks that contain variant number of branches of sub-networks, as classifiers for CIFAR-100\cite{krizhevsky2009learning}, which is a dataset for image recognition, which has 100 fine-grained classes containing 600 images each. These fine-grained classes are grouped into 20 coarse-grained classes. 

\begin{figure}[h t b p]
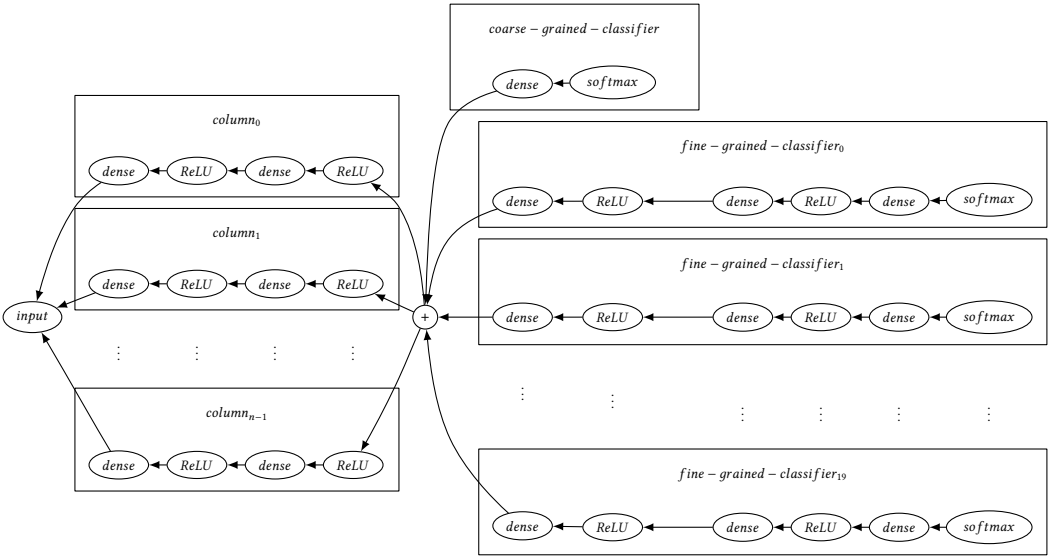

  \begin{dot2tex}[dot,mathmode,autosize,graphstyle={scale=0.54,transform shape}]
  digraph {
    rankdir=RL
    shape=circle
    overlap=scale
    nodesep=0.1
    ranksep=0.1
    rank=max

    subgraph cluster_column_0 {
      graph [ label="column_0" ]
      "fc_{0,1}" [ label=dense ]
      "fc_{0,0}" [ label=dense ]
      "ReLU_{0,1}" [ label="ReLU" ]
      "ReLU_{0,0}" [ label="ReLU" ]
      
      "ReLU_{0,1}" -> "fc_{0,1}" -> "ReLU_{0,0}" -> "fc_{0,0}"
    }
    subgraph cluster_column_1 {
      graph [ label="column_1" ]
      "fc_{1,1}" [ label=dense ]
      "fc_{1,0}" [ label=dense ]
      "ReLU_{1,1}" [ label="ReLU" ]
      "ReLU_{1,0}" [ label="ReLU" ]
      "ReLU_{1,1}" -> "fc_{1,1}" -> "ReLU_{1,0}" -> "fc_{1,0}"
    }
    
    subgraph cluster_dots {
      node [
      	shape=none 
      	label="{\vdots}" 
      ]
      graph [
      	label=""
        style=invis
      ]
      dots0 -> dots1 -> dots2 -> dots3 [ style=invis ]
   	}

    subgraph cluster_column_last {
      graph [ label="column_{n-1}" ]
      "fc_{n,1}" [ label=dense ]
      "fc_{n,0}" [ label=dense ]
      "ReLU_{n,1}" [ label="ReLU" ]
      "ReLU_{n,0}" [ label="ReLU" ]
      "ReLU_{n,1}" -> "fc_{n,1}" -> "ReLU_{n,0}" -> "fc_{n,0}"
    }
    
    subgraph cluster_coarse {
      graph [ label="coarse-grained-classifier" ]
      fc_2 [ label=dense ]
      softmax -> fc_2
    }
        
    fc_2 -> "+"

	"fc_{0,0}" -> input
    "+" -> "ReLU_{0,1}"
    
    "fc_{1,0}" -> input
    "+" -> "ReLU_{1,1}"
    
    dots3 -> input [ style=invis ]
    "+" -> dots0 [ style=invis ]
    
    "fc_{n,0}" -> input
    "+" -> "ReLU_{n,1}"
    
    subgraph cluster_fine_0 {
      graph [ label="fine-grained-classifier_0" ]
      softmax_0 [ label=softmax ]
      "fc_{0,5}" [ label=dense ]
      "fc_{0,4}" [ label=dense ]
      "fc_{0,3}" [ label=dense ]
      "ReLU_{0,4}" [ label=ReLU ]
      "ReLU_{0,3}" [ label=ReLU ]

      softmax_0 -> "fc_{0,5}" -> "ReLU_{0,4}" -> "fc_{0,4}" -> "ReLU_{0,3}" -> "fc_{0,3}"
    }
    
    subgraph cluster_fine_1 {
      graph [ label="fine-grained-classifier_1" ]
      softmax_1 [ label=softmax ]
      "fc_{1,5}" [ label=dense ]
      "fc_{1,4}" [ label=dense ]
      "fc_{1,3}" [ label=dense ]
      "ReLU_{1,4}" [ label=ReLU ]
      "ReLU_{1,3}" [ label=ReLU ]
     "softmax_1" -> "fc_{1,5}" -> "ReLU_{1,4}" -> "fc_{1,4}" -> "ReLU_{1,3}" -> "fc_{1,3}"
    }
    
    subgraph cluster_fine_dots {      
      node [
      	shape=none 
      	label="{\vdots}" 
      ]
      graph [
      	label=""
        style=invis
      ]
      "dots4" -> "dots5" -> "dots5" -> "dots6" -> "dots7" -> "dots8" -> "dots9" [ style=invis ]
    }
    
    subgraph cluster_fine_19 {
      graph [ label="fine-grained-classifier_{19}" ]
      softmax_19 [ label=softmax ]
      "fc_{19,5}" [ label=dense ]
      "fc_{19,4}" [ label=dense ]
      "fc_{19,3}" [ label=dense ]
      "ReLU_{19,4}" [ label=ReLU ]
      "ReLU_{19,3}" [ label=ReLU ]
      softmax_19 -> "fc_{19,5}" -> "ReLU_{19,4}" -> "fc_{19,4}" -> "ReLU_{19,3}" -> "fc_{19,3}"
    }

 	"fc_{0,3}" -> "+"
 	"fc_{1,3}" -> "+"
 	"dots9" -> "+" [ style=invis ]
 	"fc_{19,3}" -> "+"
  }
  \end{dot2tex}
  \caption{The models used for benchmarks}
  \label{benchmark architecture}
\end{figure}

The architecture of the network used for benchmarks summarized in Figure~\ref{benchmark architecture}.

We created $n$ columns\cite{ciregan2012multi} of experts sub-networks for feature extracting. The output from those experts are then summed for further layers and classifiers. Each column contains two dense layers
\footnote{
We use dense layers instead of convolution layers because the underlying N-dimensional array library ND4J is not able to efficiently perform immutable operation of convolution. See section~\ref{New Back-end} for discussion.
}
followed by ReLU activation layers. Those columns are independent, and we expect our framework can execute them in parallel when training or inference.

Since CIFAR-100 dataset has both coarse-grained and fine-grained labels, we created a coarse-grained classifier and 20 fine-grained classifiers. The coarse-grained classifier contains a dense layer to convert features to scores of 20 coarse-grained classes, followed by a softmax layer. Each fine-grained classifier corresponding to a coarse-grained class contains three dense layers, two ReLU activation layers, and a softmax layer, classifying 5 fine-grained classes.

We constructed mini-batches by a coarse-grained class when training. All samples in a mini-batch belongs to one coarse-grained class. With the help of the feature of dynamic neural network in DeepLearning.scala, for each training iteration, only one fine-grained classifier is used, other 19 fine-grained classifiers are skipped. When inferencing, the fine-grained classifier is chosen dynamically according to the prediction by the coarse-grained classifier. For comparison, we also created another set of benchmarks that do not skip unmatched fine-grained classifiers.

The benchmark result when training these models is shown in Table~\ref{benchmark-dont-skip-fine-grained} and Table~\ref{benchmark-skip-fine-grained} (larger score is better). The sizes of those dense layers previous to softmax layers in these models are the number of classes (20 for coarse-grained classifier, 5 for fine-grained classifiers). Other dense layers output 64 features. These models are trained in a SGD optimizer with batch size 16. Both the number of columns in each model and the number of threads used in training vary among benchmarks.

\begin{table}[htbp]
  \begin{tabular}{l|l|l|rl}
  \texttt{number of columns} & \texttt{thread pool size} & \multicolumn{2}{c}{\texttt{Score, ops/s}} \\
  \hline
  \texttt{4} & \texttt{1} & \texttt{2.550} & \scriptsize $\pm$ \texttt{0.043}  \\
  \texttt{4} & \texttt{2} & \texttt{3.233} & \scriptsize $\pm$ \texttt{0.276}  \\
  \texttt{4} & \texttt{4} & \texttt{3.338} & \scriptsize $\pm$ \texttt{0.217}  \\
  \texttt{2} & \texttt{1} & \texttt{3.345} & \scriptsize $\pm$ \texttt{0.095}  \\
  \texttt{2} & \texttt{2} & \texttt{3.987} & \scriptsize $\pm$ \texttt{0.091}  \\
  \texttt{2} & \texttt{4} & \texttt{4.828} & \scriptsize $\pm$ \texttt{0.200}  \\
  \texttt{1} & \texttt{1} & \texttt{3.854} & \scriptsize $\pm$ \texttt{0.046}  \\
  \texttt{1} & \texttt{2} & \texttt{5.239} & \scriptsize $\pm$ \texttt{0.288}  \\
  \texttt{1} & \texttt{4} & \texttt{6.328} & \scriptsize $\pm$ \texttt{0.058}  \\
  \end{tabular}
  \caption{The benchmark result when no fine-grained classifier is skipped}
  \label{benchmark-dont-skip-fine-grained}
\end{table}

\begin{table}[htbp]
  \begin{tabular}{l|l|l|l|l|rl}
  \texttt{number of columns} & \texttt{thread pool size} & \multicolumn{2}{c}{\texttt{Score, ops/s}} \\
  \hline
  \texttt{4} & \texttt{1} & \texttt{ 5.314} & \scriptsize $\pm$ \texttt{0.304}  \\
  \texttt{4} & \texttt{2} & \texttt{ 5.484} & \scriptsize $\pm$ \texttt{0.362}  \\
  \texttt{4} & \texttt{4} & \texttt{ 5.607} & \scriptsize $\pm$ \texttt{0.159}  \\
  \texttt{2} & \texttt{1} & \texttt{ 8.702} & \scriptsize $\pm$ \texttt{0.430}  \\
  \texttt{2} & \texttt{2} & \texttt{ 9.527} & \scriptsize $\pm$ \texttt{0.259}  \\
  \texttt{2} & \texttt{4} & \texttt{ 8.831} & \scriptsize $\pm$ \texttt{0.268}  \\
  \texttt{1} & \texttt{1} & \texttt{12.430} & \scriptsize $\pm$ \texttt{1.514}  \\
  \texttt{1} & \texttt{2} & \texttt{13.531} & \scriptsize $\pm$ \texttt{0.618}  \\
  \texttt{1} & \texttt{4} & \texttt{14.513} & \scriptsize $\pm$ \texttt{0.554}  \\
  \end{tabular}
  \caption{The benchmark result when unmatched fine-grained classifiers are skipped}
  \label{benchmark-skip-fine-grained}
\end{table}

The benchmark result verified the performance improvement when increasing thread pool size, since DeepLearning.scala executes independent sub-networks in parallel. This benchmark also shows a performance improvement when unmatched fine-grained classifiers are skipped.

\section{Future Works}

\subsection{New Back-end}\label{New Back-end}
Currently, DeepLearning.scala 2's built-in differentiable vector expression type \lstinline{INDArrayLayer} is based on nd4j's \lstinline{INDArray}\cite{skymind2017nd4j}.
As described in Section~\ref{training iteration}, in each training iteration, for each \gls{computational graph} node, \lstinline{forward} and \lstinline{backward} operations are performed, which internally call some methods on \lstinline{INDArray}, resulting in GPU kernel executions for nd4j's CUDA runtime. These ND4J operations have bad computational performance because:
\begin{enumerate*}
  \item ND4J is not designed for immutable tensors;\footnote{\url{https://github.com/deeplearning4j/nd4j/issues/2271\#issuecomment-344865418}}
  \item some operations\footnote{\lstinline{INDArray.broadcast} for example} are extremely slow;
  \item enqueuing a kernel is relatively expensive.
\end{enumerate*}

 We are developing a new back-end as an alternative to nd4j. The new back-end will be able to merge multiple primitive operations into one larger kernel by dynamically generating OpenCL code. The new back-end
 will support more optimized operations on the GPU and reduce the number of kernel executions. We expect our new version will achieve better computational efficiency.

\subsection{Distributed Model}

Current DeepLearning.scala is only able to run on a standalone JVM, not a distributed cluster, thus it does not support ``outrageously large neural networks''\cite{shazeer2017outrageously} that do not fit into the memory of a single node.

Since our \gls{computational graph} is made of monadic expressions that consist of closures, they can be serialized and executed remotely in theory. We are investigating how to build a distributed machine learning system based on remotely executed monadic expression. We will find out if this suggested approach can support more complex models than the parameter server approach can.

\section{Discussion}

DeepLearning.scala is an unique library among all deep learning frameworks. Our approach of AD has some attributes that never appear in previous frameworks.

\subsection{Interoperable Differentiable Computational Graph}

There were two different mechanisms in state-of-the-art deep learning frameworks:
Define-and-Run vs. Define-by-Run.

State-of-the-art Define-and-Run frameworks\cite{collobert2008torch,bergstra2010theano,jia2014caffe,chen2015mxnet,abadi2016tensorflow,intel2016bigdl,skymind2017deeplearning4j} allows users to create \glspl{computational graph}, which are immutable Abstract Syntax Trees (ASTs) of some object languages which can be evaluated by the framework runtime. Define-and-Run frameworks can schedule \glspl{computational graph} to multiple CPUs, GPUs or other devices. However, the object languages have bad interoperability with the metalanguage. For example, a DeepLearning4j user cannot use Java control flows nor call Java native methods in neural networks.

State-of-the-art Define-by-Run frameworks\cite{tokui2015chainer,neubig2017dynet,google2017eager,paszke2017pytorch} can eagerly execute actual forward pass calculation in user written code, and, at the same time, generate the internal states for running backward pass. Unlike Define-and-Run frameworks, Define-by-Run frameworks have good interoperability with the hosting language. Control flows and native function calls can be easily used during the execution of neural networks. However, Define-and-Run frameworks tend to store states and perform side effects when defining neural network structures, which makes this mechanism unsuitable for implementation in a purely functional flavor.

We discovered the third mechanism of monadic deep learning. Neural networks in DeepLearning.scala are immutable like in Define-and-Run frameworks, and interoperable with Scala like in Define-by-Run frameworks.

\subsection{AD in a Functional Library}

Reverse mode AD as a functional library was previously considered as impossible to implement without the ability to reflectively access and transform expressions associated with closures\cite{pearlmutter2008reverse}. For example, if you want to create a \lstinline{transform} function that returns the derivative for given function \lstinline{f}:

\begin{lstlisting}[float={h t b p},caption={Impossible transform function for AD}, label={transform}]
def transform(f: Double => Double): Double => Double
\end{lstlisting}

Obviously this \lstinline{transform} function is impossible without the knowledge of the implementation of \lstinline{f}.

Fortunately, in a statically typed language, the differentiable types and non-differentiable types should differ for type safety. The type signature of our AD function can use the differentiable type \lstinline{DoubleLayer} instead of \lstinline{Double}. It can be written as Listing~\ref{typeSafeTransform}:

\begin{lstlisting}[float={h t b p},caption={Type safe transform function for AD}, label={typeSafeTransform}]
def typeSafeTransform(f: Double => DoubleLayer): Double => SideEffects = { input: Double =>
  val tape = f(input).forward
  tape.backward(Do.now(1.0))
}
\end{lstlisting}

Unlike\cite{pearlmutter2008reverse}'s compiler primitive $\overleftarrow{J}$, our \lstinline{typeSafeTransform} can use the additional methods on \lstinline{DoubleLayer}. As a result, our \lstinline{typeSafeTransform} can be implemented without reflection, as an ordinary Scala function, instead of a compiler primitive or a macro.

We also change the derivative type to an opaque monadic type \lstinline{SideEffects}. Unlike numeric derivative, \lstinline{SideEffects} can contain derivatives for more than one \glspl{trainable variable}, although this change exclude the ability of higher order numerical differentiation.

\subsection{Portability}
\label{portability}

Our approach can be implemented in a library, which requires few advanced language features. Therefore, it can be ported to other languages that are not so powerful.

Our static DAGs are applicative, which can be built from normal function calls or overloaded operators without compiler-time transformation, as described in section~\ref{applicative}.

Our direct style dynamic neural networks require only one uncommon language feature, which is !-notation, or direct style monadic expression, as described in section~\ref{eager}. The usage of this feature does not prevent porting our approach to other languages, because:

\begin{itemize}
  \item The usage of !-notation is optional, as users of DeepLearning.scala can instead use \lstinline{for} comprehension if they choose not to use !-notation.
  \item Direct style monadic expressions are available in Idris (!-notation), Haskell (do-notation) and F\# (computational expressions). Our approach can be ported to those languages without a doubt.
  \item All callback functions of monadic data types, including \lstinline{Do}, \lstinline{Future} and \lstinline{UnitContinuation} will never be re-evaluated by DeepLearning.scala more than once. Therefore, those monadic data types can be replaced to generators, coroutines or one-pass continuations, which has became a widely adopted feature in many mainstream languages, including ECMAScript, TypeScript, Go, Python, C\#, C++20, Lua.
  \item For Java or other languages that do not support generators or coroutines, those !-notation can be replaced to manually written \lstinline{flatMap} calls.
\end{itemize}

\subsection{Related works}

In this section, we will discuss other deep learning frameworks in statically typed languages.

Several AD libraries\cite{bischof1992adifor,griewank1996algorithm,TapenadeRef13,baydin2015diffsharp} written in Fortran, C++ or F\# support AD via operator overloading or external preprocessor but do not scale to deep neural networks, either due to using forward mode or lacking the feature of multiple trainable variables.

Other deep learning frameworks in statically typed languages (including Scala binding of C++ frameworks)\cite{intel2016bigdl,skymind2017deeplearning4j,baydin2016hype,chen2017typesafe,zhao2017deepdsl} do not support AD, instead, they only provide their high level \gls{computational graph} APIs to compose predefined layers into neural networks. As a result, those frameworks do not have the ability to create fine-grained custom algorithms.

\cite{elliott2018simple} presented an approach that unified compiler-time translation for AD. The approach requires a compiler plug-in that translates Haskell programs into categorical form, which gives the ability to hook into each function call. By providing different instances of type classes, those function calls can translated to differential functions of different modes of AD. Unfortunately, the compiler plug-in is only available in Haskell, thus the approach is unable to be implemented in other languages. In contrast, our approach requires no metaprogramming features, thus can be implemented in many mainstream languages, as explained in section~\ref{portability}.

\cite{wangbackpropagation} discovered an approach to perform reverse mode AD in delimited continuations. Their approach is very simple: 
\begin{enumerate*}
  \item In forward pass, a chain of delimited continuations is created by \lstinline{shift} calls, which inject some hooks that contain side effects to update the derivatives.
  \item In backward pass, hooks are executed in reverse order, triggered by \lstinline{reset}.
\end{enumerate*}
However, their approach directly perform side effects in hooks, and always executes both forward pass and backward pass sequentially. In contrast, side effects in our approach are encapsulated in tasks of monadic data types, and we also provide nondeterministic implementation of applicative operations on those task types, allowing for parallel execution of both the forward pass and backward pass.

\section{Conclusion}

DeepLearning.scala is the first library that achieves all the following goals without metaprogramming technique:

\begin{itemize}
  \item static type safety
  \item purely functional interface
  \item reverse mode AD
  \item multiple \glspl{trainable variable}
  \item interoperable internal DSL
  \item dynamic neural network
  \item statically type checking
\end{itemize}

With the help of DeepLearning.scala, normal programmers are able to build complex neural networks from simple code. They still write code as usual, and the only difference is that the code written in DeepLearning.scala is differentiable, which contains \glspl{trainable variable} that learn the knowledge.

% Appendix
\appendix

\printglossary

\begin{acks}
% TODO:
\end{acks}

\clearpage

% Bibliography
\bibliography{bibliography}


\begin{thebibliography}{43}

%%% the Web address.

\ifx \showCODEN    \undefined \def \showCODEN     #1{\unskip}     \fi
\ifx \showDOI      \undefined \def \showDOI       #1{#1}\fi
\ifx \showISBNx    \undefined \def \showISBNx     #1{\unskip}     \fi
\ifx \showISBNxiii \undefined \def \showISBNxiii  #1{\unskip}     \fi
\ifx \showISSN     \undefined \def \showISSN      #1{\unskip}     \fi
\ifx \showLCCN     \undefined \def \showLCCN      #1{\unskip}     \fi
\ifx \shownote     \undefined \def \shownote      #1{#1}          \fi
\ifx \showarticletitle \undefined \def \showarticletitle #1{#1}   \fi
\ifx \showURL      \undefined \def \showURL       {\relax}        \fi
% invisible to TeX
\providecommand\bibfield[2]{#2}
\providecommand\bibinfo[2]{#2}
\providecommand\natexlab[1]{#1}
\providecommand\showeprint[2][]{arXiv:#2}

\bibitem[\protect\citeauthoryear{Abadi, Barham, Chen, Chen, Davis, Dean, Devin,
  Ghemawat, Irving, Isard, et~al\mbox{.}}{Abadi et~al\mbox{.}}{2016}]%
        {abadi2016tensorflow}
\bibfield{author}{\bibinfo{person}{Mart{\'\i}n Abadi}, \bibinfo{person}{Paul
  Barham}, \bibinfo{person}{Jianmin Chen}, \bibinfo{person}{Zhifeng Chen},
  \bibinfo{person}{Andy Davis}, \bibinfo{person}{Jeffrey Dean},
  \bibinfo{person}{Matthieu Devin}, \bibinfo{person}{Sanjay Ghemawat},
  \bibinfo{person}{Geoffrey Irving}, \bibinfo{person}{Michael Isard},
  {et~al\mbox{.}}} \bibinfo{year}{2016}\natexlab{}.
\newblock \showarticletitle{TensorFlow: A System for Large-Scale Machine
  Learning.}. In \bibinfo{booktitle}{{\em OSDI}}, Vol.~\bibinfo{volume}{16}.
  \bibinfo{pages}{265--283}.
\newblock


\bibitem[\protect\citeauthoryear{Baydin and Pearlmutter}{Baydin and
  Pearlmutter}{2016}]%
        {baydin2016hype}
\bibfield{author}{\bibinfo{person}{Atilim~Gunes Baydin} {and}
  \bibinfo{person}{Barak~A Pearlmutter}.} \bibinfo{year}{2016}\natexlab{}.
\newblock \bibinfo{title}{Hype: Compositional Machine Learning and
  Hyperparameter Optimization}.
\newblock   (\bibinfo{year}{2016}).
\newblock
\showURL{%
\url{https://hypelib.github.io/Hype/}}


\bibitem[\protect\citeauthoryear{Baydin, Pearlmutter, Radul, and
  Siskind}{Baydin et~al\mbox{.}}{2015b}]%
        {baydin2015automatic}
\bibfield{author}{\bibinfo{person}{Atilim~Gunes Baydin},
  \bibinfo{person}{Barak~A Pearlmutter}, \bibinfo{person}{Alexey~Andreyevich
  Radul}, {and} \bibinfo{person}{Jeffrey~Mark Siskind}.}
  \bibinfo{year}{2015}\natexlab{b}.
\newblock \showarticletitle{Automatic differentiation in machine learning: a
  survey}.
\newblock \bibinfo{journal}{{\em arXiv preprint arXiv:1502.05767\/}}
  (\bibinfo{year}{2015}).
\newblock


\bibitem[\protect\citeauthoryear{Baydin, Pearlmutter, and Siskind}{Baydin
  et~al\mbox{.}}{2015a}]%
        {baydin2015diffsharp}
\bibfield{author}{\bibinfo{person}{Atilim~Gunes Baydin},
  \bibinfo{person}{Barak~A Pearlmutter}, {and} \bibinfo{person}{Jeffrey~Mark
  Siskind}.} \bibinfo{year}{2015}\natexlab{a}.
\newblock \showarticletitle{Diffsharp: Automatic differentiation library}.
\newblock \bibinfo{journal}{{\em arXiv preprint arXiv:1511.07727\/}}
  (\bibinfo{year}{2015}).
\newblock


\bibitem[\protect\citeauthoryear{Bergstra, Breuleux, Bastien, Lamblin, Pascanu,
  Desjardins, Turian, Warde-Farley, and Bengio}{Bergstra et~al\mbox{.}}{2010}]%
        {bergstra2010theano}
\bibfield{author}{\bibinfo{person}{James Bergstra}, \bibinfo{person}{Olivier
  Breuleux}, \bibinfo{person}{Fr{\'e}d{\'e}ric Bastien},
  \bibinfo{person}{Pascal Lamblin}, \bibinfo{person}{Razvan Pascanu},
  \bibinfo{person}{Guillaume Desjardins}, \bibinfo{person}{Joseph Turian},
  \bibinfo{person}{David Warde-Farley}, {and} \bibinfo{person}{Yoshua Bengio}.}
  \bibinfo{year}{2010}\natexlab{}.
\newblock \showarticletitle{Theano: A CPU and GPU math compiler in Python}. In
  \bibinfo{booktitle}{{\em Proc. 9th Python in Science Conf}}.
  \bibinfo{pages}{1--7}.
\newblock


\bibitem[\protect\citeauthoryear{Bischof, Carle, Corliss, Griewank, and
  Hovland}{Bischof et~al\mbox{.}}{1992}]%
        {bischof1992adifor}
\bibfield{author}{\bibinfo{person}{Christian Bischof}, \bibinfo{person}{Alan
  Carle}, \bibinfo{person}{George Corliss}, \bibinfo{person}{Andreas Griewank},
  {and} \bibinfo{person}{Paul Hovland}.} \bibinfo{year}{1992}\natexlab{}.
\newblock \showarticletitle{ADIFOR--generating derivative codes from Fortran
  programs}.
\newblock \bibinfo{journal}{{\em Scientific Programming\/}}
  \bibinfo{volume}{1}, \bibinfo{number}{1} (\bibinfo{year}{1992}),
  \bibinfo{pages}{11--29}.
\newblock


\bibitem[\protect\citeauthoryear{Chen}{Chen}{2017}]%
        {chen2017typesafe}
\bibfield{author}{\bibinfo{person}{Tongfei Chen}.}
  \bibinfo{year}{2017}\natexlab{}.
\newblock \showarticletitle{Typesafe Abstractions for Tensor Operations}.
\newblock \bibinfo{journal}{{\em arXiv preprint arXiv:1710.06892\/}}
  (\bibinfo{year}{2017}).
\newblock


\bibitem[\protect\citeauthoryear{Chen, Li, Li, Lin, Wang, Wang, Xiao, Xu,
  Zhang, and Zhang}{Chen et~al\mbox{.}}{2015}]%
        {chen2015mxnet}
\bibfield{author}{\bibinfo{person}{Tianqi Chen}, \bibinfo{person}{Mu Li},
  \bibinfo{person}{Yutian Li}, \bibinfo{person}{Min Lin},
  \bibinfo{person}{Naiyan Wang}, \bibinfo{person}{Minjie Wang},
  \bibinfo{person}{Tianjun Xiao}, \bibinfo{person}{Bing Xu},
  \bibinfo{person}{Chiyuan Zhang}, {and} \bibinfo{person}{Zheng Zhang}.}
  \bibinfo{year}{2015}\natexlab{}.
\newblock \showarticletitle{Mxnet: A flexible and efficient machine learning
  library for heterogeneous distributed systems}.
\newblock \bibinfo{journal}{{\em arXiv preprint arXiv:1512.01274\/}}
  (\bibinfo{year}{2015}).
\newblock


\bibitem[\protect\citeauthoryear{Chollet et~al\mbox{.}}{Chollet
  et~al\mbox{.}}{2015}]%
        {chollet2015keras}
\bibfield{author}{\bibinfo{person}{Fran{\c{c}}ois Chollet} {et~al\mbox{.}}}
  \bibinfo{year}{2015}\natexlab{}.
\newblock \bibinfo{title}{Keras}.
\newblock   (\bibinfo{year}{2015}).
\newblock


\bibitem[\protect\citeauthoryear{Ciregan, Meier, and Schmidhuber}{Ciregan
  et~al\mbox{.}}{2012}]%
        {ciregan2012multi}
\bibfield{author}{\bibinfo{person}{Dan Ciregan}, \bibinfo{person}{Ueli Meier},
  {and} \bibinfo{person}{J{\"u}rgen Schmidhuber}.}
  \bibinfo{year}{2012}\natexlab{}.
\newblock \showarticletitle{Multi-column deep neural networks for image
  classification}. In \bibinfo{booktitle}{{\em Computer vision and pattern
  recognition (CVPR), 2012 IEEE conference on}}. IEEE,
  \bibinfo{pages}{3642--3649}.
\newblock


\bibitem[\protect\citeauthoryear{Collobert, Kavukcuoglu, and Farabet}{Collobert
  et~al\mbox{.}}{2008}]%
        {collobert2008torch}
\bibfield{author}{\bibinfo{person}{Ronan Collobert}, \bibinfo{person}{K
  Kavukcuoglu}, {and} \bibinfo{person}{C Farabet}.}
  \bibinfo{year}{2008}\natexlab{}.
\newblock \showarticletitle{Torch}. In \bibinfo{booktitle}{{\em Workshop on
  Machine Learning Open Source Software, NIPS}}, Vol.~\bibinfo{volume}{76}.
\newblock


\bibitem[\protect\citeauthoryear{Duchi, Hazan, and Singer}{Duchi
  et~al\mbox{.}}{2011}]%
        {duchi2011adaptive}
\bibfield{author}{\bibinfo{person}{John Duchi}, \bibinfo{person}{Elad Hazan},
  {and} \bibinfo{person}{Yoram Singer}.} \bibinfo{year}{2011}\natexlab{}.
\newblock \showarticletitle{Adaptive subgradient methods for online learning
  and stochastic optimization}.
\newblock \bibinfo{journal}{{\em Journal of Machine Learning Research\/}}
  \bibinfo{volume}{12}, \bibinfo{number}{Jul} (\bibinfo{year}{2011}),
  \bibinfo{pages}{2121--2159}.
\newblock


\bibitem[\protect\citeauthoryear{Elliott}{Elliott}{2018}]%
        {elliott2018simple}
\bibfield{author}{\bibinfo{person}{Conal Elliott}.}
  \bibinfo{year}{2018}\natexlab{}.
\newblock \showarticletitle{The simple essence of automatic differentiation}.
\newblock \bibinfo{journal}{{\em Proceedings of the ACM on Programming
  Languages\/}} \bibinfo{volume}{2}, \bibinfo{number}{ICFP}
  (\bibinfo{year}{2018}), \bibinfo{pages}{70}.
\newblock


\bibitem[\protect\citeauthoryear{Erhan, Bengio, Courville, and Vincent}{Erhan
  et~al\mbox{.}}{2009}]%
        {erhan2009visualizing}
\bibfield{author}{\bibinfo{person}{Dumitru Erhan}, \bibinfo{person}{Y Bengio},
  \bibinfo{person}{Aaron Courville}, {and} \bibinfo{person}{Pascal Vincent}.}
  \bibinfo{year}{2009}\natexlab{}.
\newblock \showarticletitle{Visualizing Higher-Layer Features of a Deep
  Network}.
\newblock  (\bibinfo{date}{01} \bibinfo{year}{2009}).
\newblock


\bibitem[\protect\citeauthoryear{Fowler}{Fowler}{2010}]%
        {fowler2010domain}
\bibfield{author}{\bibinfo{person}{Martin Fowler}.}
  \bibinfo{year}{2010}\natexlab{}.
\newblock \bibinfo{booktitle}{{\em Domain-specific languages}}.
\newblock \bibinfo{publisher}{Pearson Education}.
\newblock


\bibitem[\protect\citeauthoryear{Google Brain}{Google Brain}{2017}]%
        {google2017eager}
Google Brain \bibinfo{year}{2017}\natexlab{}.
\newblock \bibinfo{booktitle}{{\em Eager Execution: An imperative,
  define-by-run interface to TensorFlow}}.
\newblock Google Brain.
\newblock
\showURL{%
\url{https://research.googleblog.com/2017/10/eager-execution-imperative-define-by.html}}


\bibitem[\protect\citeauthoryear{Griewank, Juedes, and Utke}{Griewank
  et~al\mbox{.}}{1996}]%
        {griewank1996algorithm}
\bibfield{author}{\bibinfo{person}{Andreas Griewank}, \bibinfo{person}{David
  Juedes}, {and} \bibinfo{person}{Jean Utke}.} \bibinfo{year}{1996}\natexlab{}.
\newblock \showarticletitle{Algorithm 755: ADOL-C: a package for the automatic
  differentiation of algorithms written in C/C++}.
\newblock \bibinfo{journal}{{\em ACM Transactions on Mathematical Software
  (TOMS)\/}} \bibinfo{volume}{22}, \bibinfo{number}{2} (\bibinfo{year}{1996}),
  \bibinfo{pages}{131--167}.
\newblock


\bibitem[\protect\citeauthoryear{Gurnell}{Gurnell}{2017}]%
        {gurnelltype}
\bibfield{author}{\bibinfo{person}{Dave Gurnell}.}
  \bibinfo{year}{2017}\natexlab{}.
\newblock \bibinfo{booktitle}{{\em The Type Astronaut's Guide to Shapeless}}.
\newblock \bibinfo{publisher}{Underscore Consulting LLP}.
\newblock
\showISBNx{9781365613524}
\showURL{%
\url{https://underscore.io/books/shapeless-guide/}}


\bibitem[\protect\citeauthoryear{Hasco{\"e}t and Pascual}{Hasco{\"e}t and
  Pascual}{2013}]%
        {TapenadeRef13}
\bibfield{author}{\bibinfo{person}{L. Hasco{\"e}t} {and} \bibinfo{person}{V.
  Pascual}.} \bibinfo{year}{2013}\natexlab{}.
\newblock \showarticletitle{The {T}apenade {A}utomatic {D}ifferentiation tool:
  {P}rinciples, {M}odel, and {S}pecification}.
\newblock \bibinfo{journal}{{\em {ACM} {T}ransactions {O}n {M}athematical
  {S}oftware\/}} \bibinfo{volume}{39}, \bibinfo{number}{3}
  (\bibinfo{year}{2013}).
\newblock
\showURL{%
\url{http://dx.doi.org/10.1145/2450153.2450158}}


\bibitem[\protect\citeauthoryear{Iandola, Moskewicz, Karayev, Girshick,
  Darrell, and Keutzer}{Iandola et~al\mbox{.}}{2014}]%
        {iandola2014densenet}
\bibfield{author}{\bibinfo{person}{Forrest Iandola}, \bibinfo{person}{Matt
  Moskewicz}, \bibinfo{person}{Sergey Karayev}, \bibinfo{person}{Ross
  Girshick}, \bibinfo{person}{Trevor Darrell}, {and} \bibinfo{person}{Kurt
  Keutzer}.} \bibinfo{year}{2014}\natexlab{}.
\newblock \showarticletitle{Densenet: Implementing efficient convnet descriptor
  pyramids}.
\newblock \bibinfo{journal}{{\em arXiv preprint arXiv:1404.1869\/}}
  (\bibinfo{year}{2014}).
\newblock


\bibitem[\protect\citeauthoryear{Intel}{Intel}{2016}]%
        {intel2016bigdl}
\bibfield{author}{\bibinfo{person}{Intel}.} \bibinfo{year}{2016}\natexlab{}.
\newblock \bibinfo{title}{BigDL}.
\newblock   (\bibinfo{year}{2016}).
\newblock
\showURL{%
\url{https://github.com/intel-analytics/BigDL}}


\bibitem[\protect\citeauthoryear{Jia, Shelhamer, Donahue, Karayev, Long,
  Girshick, Guadarrama, and Darrell}{Jia et~al\mbox{.}}{2014}]%
        {jia2014caffe}
\bibfield{author}{\bibinfo{person}{Yangqing Jia}, \bibinfo{person}{Evan
  Shelhamer}, \bibinfo{person}{Jeff Donahue}, \bibinfo{person}{Sergey Karayev},
  \bibinfo{person}{Jonathan Long}, \bibinfo{person}{Ross Girshick},
  \bibinfo{person}{Sergio Guadarrama}, {and} \bibinfo{person}{Trevor Darrell}.}
  \bibinfo{year}{2014}\natexlab{}.
\newblock \showarticletitle{Caffe: Convolutional architecture for fast feature
  embedding}. In \bibinfo{booktitle}{{\em Proceedings of the 22nd ACM
  international conference on Multimedia}}. ACM, \bibinfo{pages}{675--678}.
\newblock


\bibitem[\protect\citeauthoryear{Kingma and Ba}{Kingma and Ba}{2014}]%
        {kingma2014adam}
\bibfield{author}{\bibinfo{person}{Diederik Kingma} {and}
  \bibinfo{person}{Jimmy Ba}.} \bibinfo{year}{2014}\natexlab{}.
\newblock \showarticletitle{Adam: A method for stochastic optimization}.
\newblock \bibinfo{journal}{{\em arXiv preprint arXiv:1412.6980\/}}
  (\bibinfo{year}{2014}).
\newblock


\bibitem[\protect\citeauthoryear{Krizhevsky and Hinton}{Krizhevsky and
  Hinton}{2009}]%
        {krizhevsky2009learning}
\bibfield{author}{\bibinfo{person}{Alex Krizhevsky} {and}
  \bibinfo{person}{Geoffrey Hinton}.} \bibinfo{year}{2009}\natexlab{}.
\newblock \showarticletitle{Learning multiple layers of features from tiny
  images}.
\newblock  (\bibinfo{year}{2009}).
\newblock


\bibitem[\protect\citeauthoryear{Liu and Deng}{Liu and Deng}{2017}]%
        {liu2017dynamic}
\bibfield{author}{\bibinfo{person}{Lanlan Liu} {and} \bibinfo{person}{Jia
  Deng}.} \bibinfo{year}{2017}\natexlab{}.
\newblock \showarticletitle{Dynamic deep neural networks: Optimizing
  accuracy-efficiency trade-offs by selective execution}.
\newblock \bibinfo{journal}{{\em arXiv preprint arXiv:1701.00299\/}}
  (\bibinfo{year}{2017}).
\newblock


\bibitem[\protect\citeauthoryear{McBride and Paterson}{McBride and
  Paterson}{2008}]%
        {mcbride2008applicative}
\bibfield{author}{\bibinfo{person}{Conor McBride} {and} \bibinfo{person}{Ross
  Paterson}.} \bibinfo{year}{2008}\natexlab{}.
\newblock \showarticletitle{Applicative programming with effects}.
\newblock \bibinfo{journal}{{\em Journal of functional programming\/}}
  \bibinfo{volume}{18}, \bibinfo{number}{1} (\bibinfo{year}{2008}),
  \bibinfo{pages}{1--13}.
\newblock


\bibitem[\protect\citeauthoryear{Neubig, Dyer, Goldberg, Matthews, Ammar,
  Anastasopoulos, Ballesteros, Chiang, Clothiaux, Cohn, et~al\mbox{.}}{Neubig
  et~al\mbox{.}}{2017}]%
        {neubig2017dynet}
\bibfield{author}{\bibinfo{person}{Graham Neubig}, \bibinfo{person}{Chris
  Dyer}, \bibinfo{person}{Yoav Goldberg}, \bibinfo{person}{Austin Matthews},
  \bibinfo{person}{Waleed Ammar}, \bibinfo{person}{Antonios Anastasopoulos},
  \bibinfo{person}{Miguel Ballesteros}, \bibinfo{person}{David Chiang},
  \bibinfo{person}{Daniel Clothiaux}, \bibinfo{person}{Trevor Cohn},
  {et~al\mbox{.}}} \bibinfo{year}{2017}\natexlab{}.
\newblock \showarticletitle{DyNet: The Dynamic Neural Network Toolkit}.
\newblock \bibinfo{journal}{{\em arXiv preprint arXiv:1701.03980\/}}
  (\bibinfo{year}{2017}).
\newblock


\bibitem[\protect\citeauthoryear{Oliveira, Moors, and Odersky}{Oliveira
  et~al\mbox{.}}{2010}]%
        {oliveira2010type}
\bibfield{author}{\bibinfo{person}{BCDS Oliveira}, \bibinfo{person}{A Moors},
  {and} \bibinfo{person}{M Odersky}.} \bibinfo{year}{2010}\natexlab{}.
\newblock \showarticletitle{Type classes as objects and implicits}.
\newblock \bibinfo{journal}{{\em ACM SIGPLAN Notices\/}}
  (\bibinfo{year}{2010}).
\newblock


\bibitem[\protect\citeauthoryear{Osheim and Cantero}{Osheim and
  Cantero}{2017}]%
        {erik2017opaque}
\bibfield{author}{\bibinfo{person}{Erik Osheim} {and}
  \bibinfo{person}{Jorge~Vicente Cantero}.} \bibinfo{year}{2017}\natexlab{}.
\newblock \bibinfo{booktitle}{{\em SIP-35 - Opaque types}}.
\newblock
\showURL{%
\url{https://docs.scala-lang.org/sips/opaque-types.html}}


\bibitem[\protect\citeauthoryear{Paszke, Gross, Chintala, and Chanan}{Paszke
  et~al\mbox{.}}{2017}]%
        {paszke2017pytorch}
\bibfield{author}{\bibinfo{person}{Adam Paszke}, \bibinfo{person}{Sam Gross},
  \bibinfo{person}{Soumith Chintala}, {and} \bibinfo{person}{Gregory Chanan}.}
  \bibinfo{year}{2017}\natexlab{}.
\newblock \bibinfo{booktitle}{{\em PyTorch: Tensors and Dynamic neural networks
  in Python with strong GPU acceleration}}.
\newblock
\showURL{%
\url{http://pytorch.org/}}


\bibitem[\protect\citeauthoryear{Pearlmutter and Siskind}{Pearlmutter and
  Siskind}{2008}]%
        {pearlmutter2008reverse}
\bibfield{author}{\bibinfo{person}{Barak~A Pearlmutter} {and}
  \bibinfo{person}{Jeffrey~Mark Siskind}.} \bibinfo{year}{2008}\natexlab{}.
\newblock \showarticletitle{Reverse-mode AD in a functional framework: Lambda
  the ultimate backpropagator}.
\newblock \bibinfo{journal}{{\em ACM Transactions on Programming Languages and
  Systems (TOPLAS)\/}} \bibinfo{volume}{30}, \bibinfo{number}{2}
  (\bibinfo{year}{2008}), \bibinfo{pages}{7}.
\newblock


\bibitem[\protect\citeauthoryear{Rumelhart, Hinton, and Williams}{Rumelhart
  et~al\mbox{.}}{1985}]%
        {rumelhart1985learning}
\bibfield{author}{\bibinfo{person}{David~E Rumelhart},
  \bibinfo{person}{Geoffrey~E Hinton}, {and} \bibinfo{person}{Ronald~J
  Williams}.} \bibinfo{year}{1985}\natexlab{}.
\newblock \bibinfo{booktitle}{{\em Learning internal representations by error
  propagation}}.
\newblock \bibinfo{type}{{T}echnical {R}eport}.
  \bibinfo{institution}{California Univ San Diego La Jolla Inst for Cognitive
  Science}.
\newblock


\bibitem[\protect\citeauthoryear{Sabry and Felleisen}{Sabry and
  Felleisen}{1993}]%
        {sabry1993reasoning}
\bibfield{author}{\bibinfo{person}{Amr Sabry} {and} \bibinfo{person}{Matthias
  Felleisen}.} \bibinfo{year}{1993}\natexlab{}.
\newblock \showarticletitle{Reasoning about programs in continuation-passing
  style}.
\newblock \bibinfo{journal}{{\em Lisp and symbolic computation\/}}
  \bibinfo{volume}{6}, \bibinfo{number}{3-4} (\bibinfo{year}{1993}),
  \bibinfo{pages}{289--360}.
\newblock


\bibitem[\protect\citeauthoryear{Shazeer, Mirhoseini, Maziarz, Davis, Le,
  Hinton, and Dean}{Shazeer et~al\mbox{.}}{2017}]%
        {shazeer2017outrageously}
\bibfield{author}{\bibinfo{person}{Noam Shazeer}, \bibinfo{person}{Azalia
  Mirhoseini}, \bibinfo{person}{Krzysztof Maziarz}, \bibinfo{person}{Andy
  Davis}, \bibinfo{person}{Quoc Le}, \bibinfo{person}{Geoffrey Hinton}, {and}
  \bibinfo{person}{Jeff Dean}.} \bibinfo{year}{2017}\natexlab{}.
\newblock \showarticletitle{Outrageously large neural networks: The
  sparsely-gated mixture-of-experts layer}.
\newblock \bibinfo{journal}{{\em arXiv preprint arXiv:1701.06538\/}}
  (\bibinfo{year}{2017}).
\newblock


\bibitem[\protect\citeauthoryear{Shipilev}{Shipilev}{2018}]%
        {shipilevjmh}
\bibfield{author}{\bibinfo{person}{Aleksey Shipilev}.}
  \bibinfo{year}{2018}\natexlab{}.
\newblock \bibinfo{title}{JMH: Java Microbenchmark Harness}.
\newblock   (\bibinfo{year}{2018}).
\newblock
\showURL{%
\url{http://openjdk.java.net/projects/code-tools/jmh/}}


\bibitem[\protect\citeauthoryear{Skymind}{Skymind}{2017a}]%
        {skymind2017deeplearning4j}
Skymind \bibinfo{year}{2017}\natexlab{a}.
\newblock \bibinfo{booktitle}{{\em Deeplearning4j: Open-Source, Distributed,
  Deep Learning Library for the JVM}}.
\newblock Skymind.
\newblock
\showURL{%
\url{https://deeplearning4j.org/}}


\bibitem[\protect\citeauthoryear{Skymind}{Skymind}{2017b}]%
        {skymind2017nd4j}
Skymind \bibinfo{year}{2017}\natexlab{b}.
\newblock \bibinfo{booktitle}{{\em ND4J: N-Dimensional Arrays for Java}}.
\newblock Skymind.
\newblock
\showURL{%
\url{https://nd4j.org/}}


\bibitem[\protect\citeauthoryear{Tokui, Oono, Hido, and Clayton}{Tokui
  et~al\mbox{.}}{2015}]%
        {tokui2015chainer}
\bibfield{author}{\bibinfo{person}{Seiya Tokui}, \bibinfo{person}{Kenta Oono},
  \bibinfo{person}{Shohei Hido}, {and} \bibinfo{person}{Justin Clayton}.}
  \bibinfo{year}{2015}\natexlab{}.
\newblock \showarticletitle{Chainer: a next-generation open source framework
  for deep learning}. In \bibinfo{booktitle}{{\em Proceedings of workshop on
  machine learning systems (LearningSys) in the twenty-ninth annual conference
  on neural information processing systems (NIPS)}}, Vol.~\bibinfo{volume}{5}.
\newblock


\bibitem[\protect\citeauthoryear{Wang, Decker, Wu, Essertel, and Rompf}{Wang
  et~al\mbox{.}}{[n. d.]}]%
        {wangbackpropagation}
\bibfield{author}{\bibinfo{person}{Fei Wang}, \bibinfo{person}{James Decker},
  \bibinfo{person}{Xilun Wu}, \bibinfo{person}{Gr{\'e}gory Essertel}, {and}
  \bibinfo{person}{Tiark Rompf}.} \bibinfo{year}{[n. d.]}\natexlab{}.
\newblock \showarticletitle{Backpropagation with Continuation Callbacks:
  Foundations for Efficient and Expressive Differentiable Programming}.
\newblock  (\bibinfo{year}{[n. d.]}).
\newblock


\bibitem[\protect\citeauthoryear{Yang}{Yang}{2017}]%
        {yang2017dsl}
\bibfield{author}{\bibinfo{person}{Bo Yang}.} \bibinfo{year}{2017}\natexlab{}.
\newblock \bibinfo{title}{Dsl.scala: a framework to create DSL in Scala}.
  (\bibinfo{year}{2017}).
\newblock
\showURL{%
\url{https://github.com/ThoughtWorksInc/Dsl.scala/}}


\bibitem[\protect\citeauthoryear{Yoshida}{Yoshida}{2017}]%
        {kenji2017scalaz}
\bibfield{author}{\bibinfo{person}{Kenji Yoshida}.}
  \bibinfo{year}{2017}\natexlab{}.
\newblock \bibinfo{booktitle}{{\em Scalaz: An extension to the core scala
  library}}.
\newblock
\showURL{%
\url{https://scalaz.github.io/scalaz/}}


\bibitem[\protect\citeauthoryear{Zeiler}{Zeiler}{2012}]%
        {zeiler2012adadelta}
\bibfield{author}{\bibinfo{person}{Matthew~D Zeiler}.}
  \bibinfo{year}{2012}\natexlab{}.
\newblock \showarticletitle{ADADELTA: an adaptive learning rate method}.
\newblock \bibinfo{journal}{{\em arXiv preprint arXiv:1212.5701\/}}
  (\bibinfo{year}{2012}).
\newblock


\bibitem[\protect\citeauthoryear{Zhao, Huang, and Cao}{Zhao
  et~al\mbox{.}}{2017}]%
        {zhao2017deepdsl}
\bibfield{author}{\bibinfo{person}{Tian Zhao}, \bibinfo{person}{Xiaobing
  Huang}, {and} \bibinfo{person}{Yu Cao}.} \bibinfo{year}{2017}\natexlab{}.
\newblock \showarticletitle{DeepDSL: A Compilation-based Domain-Specific
  Language for Deep Learning}.
\newblock \bibinfo{journal}{{\em arXiv preprint arXiv:1701.02284\/}}
  (\bibinfo{year}{2017}).
\newblock


\end{thebibliography}

\end{document}